\documentclass[12pt]{article}

\usepackage{latexsym,amsmath,amssymb,theorem,epsfig} 

\DeclareFontFamily{OT1}{rsfs10}{} 
\DeclareFontShape{OT1}{rsfs10}{m}{n}{ <-> rsfs10 }{} 
\DeclareMathAlphabet{\mathscript}{OT1}{rsfs10}{m}{n} 

\numberwithin{equation}{section}

\newcommand{\cp}[1]{{\mathbb C}{\mathbb P}^{#1}}

\newcommand{\ns}{\normalsize}

\newcommand{\bX}{{\mathbf X}}
\newcommand{\bY}{{\mathbf Y}}
\newcommand{\bZ}{{\mathbf Z}}

\def\cC{{\mathcal C}}
\def\cE{{\mathcal E}}
\def\cF{{\mathcal F}}

\def\cL{{\mathcal L}}

\def\cN{{\mathcal N}}
\def\cO{{\mathcal O}}

\def\cS{{\mathcal S}}

\theoremstyle{plain} 

{\theorembodyfont{\rmfamily} }

\begin{document}


\begin{titlepage}

\vspace{-5cm}

\title{
   \hfill{\ns UPR-979T} \\[1em]
   {\LARGE Vector Bundle Moduli and Small Instanton Transitions} \\[1em] } 
\author{
   Evgeny Buchbinder$^1$, Ron Donagi$^2$
   and Burt A. Ovrut$^1$\\[0.5em]
   {\ns $^1$Department of Physics, University of Pennsylvania} \\[-0.4em]
   {\ns Philadelphia, PA 19104--6396}\\
   {\ns $^2$Department of Mathematics, University of Pennsylvania} 
   \\[-0.4em]
   {\ns Philadelphia, PA 19104--6395, USA}\\} 
   
\date{}

\maketitle

\begin{abstract}

We give the general presciption for calculating the moduli of 
irreducible, stable $SU(n)$ holomorphic vector bundles with positive spectral
covers over elliptically fibered Calabi-Yau threefolds. Explicit results 
are presented for Hirzebruch
base surfaces $B={\mathbb F}_{r}$. The transition moduli that are produced
by
chirality changing small instanton phase transitions are defined and
specifically enumerated. The origin of these moduli, as the deformations
of
the spectral cover restricted to the ``lift'' of the horizontal curve of
the 
$M$5-brane, is discussed. We present an alternative description of the
transition moduli as the sections of rank $n$ holomorphic vector bundles 
over the $M$5-brane curve and
give explicit examples. Vector bundle moduli appear 
as gauge singlet scalar fields in the
effective low-energy actions of heterotic superstrings and heterotic
$M$-theory. 
 
\end{abstract}

\thispagestyle{empty}

\end{titlepage}


\section{Introduction:}


$G$-instantons, that is, solutions of the Hermitian Yang-Mills equations
with
structure group $G \subset E_{8}$ on complex manifolds, play a prominent 
role in determining the
vacuum structure of both the weakly coupled heterotic superstring 
\cite{CHSW} (see also \cite{GSW}) 
and
heterotic $M$-theory \cite{hw1,ssea,losw1}. Unfortunately, for manifolds
of dimension
greater than one, it has proven difficult or impossible to find the
explicit
one-form gauge connections that solve these equations. However, it was
demonstrated by Donaldson \cite{donald} and 
Uhlenbeck and Yau \cite{yau} that there is a
one-to-one relationship between the solutions of the Hermitian Yang-Mills
equations and the construction of stable, holomorphic vector bundles with
structure group $G$ over the same complex manifold. Happily, construction
of
such vector bundles is relatively straightforward, as was demonstrated for
elliptically fibered Calabi-Yau manifolds by Friedman, Morgan and Witten
\cite{FMW1} and Donagi \cite{AJ}. Hence, to construct 
vacua of heterotic superstrings and
$M$-theory, one can either attempt to solve the Hermitian Yang-Mills
equations or, more simply, construct the associated stable, holomorphic
vector
bundles. This latter approach has been used extensively in 
\cite{curio, andreas, di} to study charged matter in heterotic 
compactifications
and in \cite{gub, gub2,smb} to search for
phenomenologically relevant grand unified or standard model vacua in
heterotic
$M$-theory.

Any solution of the Hermitian Yang-Mills equations will depend on a number
of
independent integration constants, the ``instanton moduli''. The number
and 
structure of these moduli will, in turn, depend on the geometry of the 
complex manifold, on the chosen structure group of the instanton and on
the 
type of solution itself. When heterotic superstrings or $M$-theory are 
compactified to lower dimension on a complex manifold, instanton 
moduli will appear, not as constants, but as independent, gauge singlet 
scalar fields on the uncompactified space. These moduli fields, although
relatively little explored, play a substantial role in the physics of both
heterotic string theory and heterotic $M$-theory. First, we note that
superstrings wrapped on holomorphic curves in the complex manifold will
induce a non-perturbative contribution to the superpotential of many of
the
compactification moduli, including the 
instanton moduli \cite{Becker,Dinst}. 
Computations of the superpotential for both Calabi-Yau and $M5$-brane
moduli
have been presented in \cite{lima1,lima2,moore1,moore2,ckrause}. 
We will give a detailed calculation of the
superpotential for instanton moduli in a subsequent 
publication \cite{future}. Suffice it
here to say that this superpotential mixes instanton moduli with other
geometric moduli. It follows that including instanton moduli, and
computing their
superpotential, is essential for the over-all study of the stability of
heterotic superstring and $M$-theory vacua. Secondly, instanton moduli
also
play an important role in phenomenology and 
cosmology \cite{low4,hos,blo}. For example, in the
recently proposed Ekpyrotic scenarios of the early universe
\cite{stok,stokd,stoks,stok2,pn1},
$M$-brane moduli interact with and transfer energy to instanton moduli,
both
before, during and after the cataclysmic collision that produces the ``big
bang''.
Hence, much of the structure of primordial fluctuations, reheating and
the value of the cosmological constant, for example, depend on the theory
of
the instanton moduli.

As stated above, it is generically unknown how to compute explicit
solutions of the Hermitian Yang-Mills equations on complex manifolds. 
Hence, the number and properties of instanton moduli are difficult to
determine from this point of view. However, the theorems of Donaldson,
Uhlenbeck and Yau allow one to approach this issue from the point of view
of
the associated stable, holomorphic vector bundles. Here, one is much 
more successful. Specifically, we will show,
within
the context of irreducible, stable, holomorphic $SU(n)$ vector
bundles with positive spectral covers
over elliptically fibered Calabi-Yau threefolds, how to compute the the
precise
number of ``vector bundle moduli'' and, in the process, exhibit their
origin
and some of their properties. These vector bundle moduli are, of course,
identical to the instanton moduli of the associated Yang-Mills connection.
We
will, henceforth, refer to these moduli, in any context, as vector bundle
moduli.

We can, in fact, accomplish much more than the enumeration of the vector
bundle moduli. As discussed in \cite{tony}, in heterotic $M$-theory a bulk
space
$M$5-brane, wrapped on a holomorphic curve in the Calabi-Yau threefold,
can
collide with an ``end-of-the-world'' boundary brane. This collision, 
called a ``small instanton'' phase transition, modifies the smooth vector
bundle
$V$ on the boundary brane, first to a singular ``torsion free sheaf'' 
(the small instanton) and then to another smooth vector bundle $V'$, 
generically of a different topological type 
than $V$. During this process, the number of vector
bundle
moduli changes. In this paper, we will compute, within the class of
chirality altering small instanton transitions, the specific change in the
number of
vector bundle moduli. The new moduli that appear during the collision are
termed 
``transition moduli'', and we will elucidate their origin and structure in
detail. These transition moduli, in interaction with the $M$-brane moduli,
determine
much of the dynamics of small instanton phase transitions. For example,
their
vacuum values control the strength of non-perturbative corrections to the
superpotential. In a cosmological context, the transition moduli will be 
responsible for processes such as reheating, the cosmological constant and
the
``bounce'' dynamics in Ekpyrotic theories of the early universe.

Specifically, in this paper we will do the following. 
In Section 2, we discuss the explicit relationship between the instanton
moduli of an exact solution of the Hermitian Yang-Mills equations 
and the vector bundle
moduli of the associated stable, holomorphic vector bundle. In the simple
context of compactification on a complex, one-dimensional torus, we compute the 
number of moduli and their moduli space both
analytically, in the Hermitian Yang-Mills approach, and algebraically, from the
vector bundle. We show that we get the same answer either way. Unfortunately, 
this is essentially the only
case where the analytic moduli can be completely described. This motivates our
approach in the rest of the paper, where we will enumerate 
and describe the properties 
of these moduli within the context of stable, holomorphic vector bundles.
In Section 3, we present a
brief review of the theory of irreducible, stable, holomorphic
$SU(n)$
vector bundles over elliptically fibered Calabi-Yau threefolds. We also define
the concept of a positive spectral cover, and give the criteria 
for the positivity of spectral covers over threefolds with base $B={\mathbb F}_{r}$.
An explict
discussion of the origin and the computation of the number of vector
bundle moduli
in this context is given in Section 4. Section 5 is devoted to moduli in
chirality changing small instanton phase transitions. First, we briefly
discuss the basic concepts of these transitions and then explicitly compute the
number of transition moduli. 
We then show that these moduli can be localized on a surface within the
Calabi-Yau threefold. This surface is the ``lift'' of the
$M$5-brane curve. Specifically, we identify the transition moduli with
the deformations of the restriction of the spectral cover to this surface. In
Section 6, we give a second interpretation of the transition moduli by
evaluating them on the $M$5-brane curve itself. In this context, 
they appear as the holomorphic 
sections of a certain rank $n$ vector bundle on the $M$5-brane curve. 
We calculate this vector bundle explicitly.

This is the first in a series of papers that will discuss the properties
of vector
bundle moduli. The present paper is, necessarily, somewhat mathematical in
its
content. This reflects that fact that one must work directly with
holomorphic
vector bundles, since solutions of the associated Hermitian Yang-Mills
equations are unknown. Be that as it may, the theory of vector bundle
moduli
potentially has considerable physical importance. We will elucidate the more
physical
aspects of these moduli in future publications.  


\section{Vector Bundle Moduli on a Torus:}


Consider an $N=1$ supersymmetric $SU(m)$ Yang-Mills theory in flat
six-dimensional space. The action is given by
\begin{equation}
S_{YM}= -\frac{1}{4g_6^2} \int {\rm d}^6 x 
{\rm tr} F_{MN}^2 + \cdots ,
\label{eq:B1}
\end{equation}
where $M,N=0,1,\dots,5$, $g_{6}$ is the six-dimensional coupling parameter
and
the dots stand for terms containing fermions. Now, compactify this theory
on a
torus, $T^{2}$. To do this, begin by splitting the coordinates $x^{M}$
into
$x^{\mu}$, $\mu=0,1,\dots,3$ parameterizing four-dimensional Minkowski
space,
${\mathbb R}^{4}$, and complex coordinates $z$,$\bar{z}$ on $T^{2}$.
Similarly,
the $SU(m)$ Lie algebra valued connection $A_{M}$ decomposes into the
four-dimensional gauge connection $A_{\mu}$ on ${\mathbb R}^{4}$ and two
scalars $A_{z}$ and $A_{\bar{z}}(=A_{z}^{\dagger})$. We now want to search
for
solutions of the Yang-Mills equations restricted to $T^{2}$ that are Lie
algebra valued in a subgroup $SU(n) \subset SU(m)$ and preserve $N=1$
supersymmetry on ${\mathbb R}^{4}$. For this to be
the case, $A_{z}$ and $A_{\bar{z}}$ must satisfy the Hermitian Yang-Mills
equations which, on $T^{2}$, are simply the zero curvature equation
\begin{equation}
F_{z \bar z}=0,
\label{eq:B2}
\end{equation}
where 
\begin{equation}
F_{z \bar z}= \partial_z A_{\bar z} - 
\partial_{\bar z} A_z + [A_z, A_{\bar z}].
\label{eq:B3}
\end{equation}
That is, $A_{z}$ is a flat connection Lie algebra valued in $SU(n)$. 
If we express all quantities in terms of the real coordinates $y_{1}$,$y_{2}$
on $T^{2}$, then the connection on the torus is written as $A_{i}$ for $i=1,2$.
The most general solution of~\eqref{eq:B2} is well-known and given by the
Wilson line
\begin{equation}
A_i = {\rm diag} (A_i^1, \dots A_i^n), \qquad \sum_{k=1}^{n}A_i^k =0,
\label{eq:B9} 
\end{equation}
where
\begin{equation}
A_i^k = C_i^k + \frac{n_i^k}{R_i}, 
\label{eq:B10}
\end{equation}
$R_1$,$R_2$ are the radii of the torus $T^{2}$ and 
$n_{1}^{k}=l_{k}$, $n_{2}^{k}=m_{k}$ where $l_k$, $m_k$ 
for $k=1, \dots, n $ are any integers satisfying 
\begin{equation}
\sum_{k=1}^{n} l_k = 0, \qquad \sum_{k=1}^{n} m_k = 0. 
\label{eq:B8}
\end{equation}
The parameters $C^{k}_{i}$ for $i=1,2$ are real numbers
subject to the constraints
\begin{equation}
\sum_{k=1}^{n} C_{i}^k=0
\label{eq:B6.555}
\end{equation}
Hence, only $2(n-1)$ of the parameters $C^{k}_{i}$ are independent.
We conclude, therefore,
that
the zero curvature equation~\eqref{eq:B2} has constant
solutions~\eqref{eq:B9},~\eqref{eq:B10} on $T^{2}$ that are not gauge
equivalent to zero. They are, however, periodic with periods $R_{1}$ and
$R_{2}$ in the two compact torus directions. These $2(n-1)$ real constants
are
the ``instanton moduli'' of this gauge configuration. It follows that the
moduli space of flat, $SU(n)$ valued Yang-Mills gauge connections has real
dimension $2(n-1)$ and is given by
\begin{equation}
{\cal M}_{flat}(SU(n)) \simeq (T^2)^{n-1}/ \Sigma_n,
\label{eq:B11}
\end{equation}
where $\Sigma_n$ denotes the group of permutations of $A_i^k$. This space
can
be shown to be equivalent to the $n-1$ dimensional complex moduli space
\begin{equation}
{\cal M}_{flat}(SU(n)) \simeq {\mathbb C}{\mathbb P}^{n-1}.
\label{eq:B12}
\end{equation}

Having found the solution~\eqref{eq:B9} of the zero curvature equation for
$SU(n)$ Lie algebra valued gauge connections on a simple torus, we see
that in
the compactification proceedure these moduli will appear as functions,
$A_{i}^{k}(x^{\mu})$, of the coordinates $x^{\mu}$,
$\mu=0,1,\dots,3$ on ${\mathbb R}^{4}$. The solution~\eqref{eq:B9} is
interpreted as the vacuum expectation values
\begin{equation}
C_{i}^{k}=<A_{i}^{k}(x^{\mu})>.
\label{eq:B12a}
\end{equation}
These expectation values break the gauge group $SU(m)$ down to $U(m-n)$,
which is the commutant of $SU(n)$ in $SU(m)$. The fluctuations
\begin{equation}
\phi_{i}^{k}(x^{\mu})=A_{i}^{k}(x^{\mu})-C_{i}^{k}
\label{eq:B12b}
\end{equation}
will appear as massless scalar fields in the four-dimensional effective
action
and are singlets under $U(m-n)$. Returning to complex coordinates
$z$,$\bar{z}$ on $T^{2}$, these scalars 
\begin{equation}
\phi_{z}^{k}=\sqrt{2}(\phi_{1}^{k}+i\phi_{2}^{k}), \qquad
\sum_{k=1}^{n}\phi_{z}^{k}=0
\label{eq:B12c}
\end{equation}
form the lowest components of $n-1$, $N=1$ chiral supermultiplets. The
bosonic
part of the four-dimensional $N=1$ supersymmentric effective action takes
the
form
\begin{equation}
S_{YM}= -\frac{1}{4g_4^2} \int {\rm d}^4 x 
({\rm tr} F_{\mu \nu}^2 + 4 \sum_{k=1}^{n}
\partial_{\mu} \phi_{\bar z}^{k} \partial^{\mu}\phi_z^{k} \cdots ), 
\label{eq:B13}
\end{equation}
where $g_4$ is the four-dimensional coupling parameter, $F_{\mu \nu}$ is
the
bosonic field strength of the $U(m-n)$ Lie algebra valued super
Yang-Mills
connection and $\phi_{\bar z}^{k}= \phi_z^{k*}$. There are, of course,
also
chiral supermultiplets charged under $U(m-n)$ indicated in the
action~\eqref{eq:B13} by the dots. We conclude that, for the simple case
of
the torus $T^{2}$, one can completely solve the Hermitian Yang-Mills
equations
and determine the instanton moduli. Be that as it may, it is of interest
to
see whether one could arrive at the same information by analyzing the
associated holomorphic vector bundle. We turn to this now.

It is well-known that every stable $U(n)$ holomorphic vector bundle $V$ on
$T^{2}$ is a sum of $n$ line bundles, which we denote by $L_{i}$, each
with
vanishing first Chern class. That is
\begin{equation}
V= \oplus_{i=1}^{n} L_i \qquad c_1(L_i)=0.
\label{eq:B14}
\end{equation}
The additional restriction to structure group $SU(n)$ implies that the
product
of all the line bundles $L_{i}$ is the trivial bundle
\begin{equation}
\otimes_{i=1}^{n} L_i = \cO.
\label{eq:B15}
\end{equation}
On the torus, the fact that a line bundle has vanishing first Chern class
implies that it has a meromorphic section with exactly one zero and one
pole.
Denote the zero of the meromorphic section of line bundle $L_{i}$ by
$Q_{i}$.
The pole can be chosen to be the same point for all line bundles $L_{i}$
and
will be denoted by $P$. The divisor $Q_{i}-P$ has zero degree and,
therefore,
the corresponding line bundle $\cO (Q_i) \otimes \cO (P)^{-1}$ has
vanishing
first Chern class. Clearly
\begin{equation}
L_i \simeq \cO (Q_i) \otimes \cO (P)^{-1}.
\label{eq:B16} 
\end{equation}
Therefore, line bundle $L_{i}$ is uniquely determined by the point $Q_{i}$
and
the deformation of this bundle specified by how this point can move in the
torus. That is, the moduli space of the line bundle $L_{i}$ is isomorphic
to
the torus itself. Note that using~\eqref{eq:B16}, the $SU(n)$ 
condition~\eqref{eq:B15} becomes
\begin{equation}
\sum_{i=1}^{n}(Q_{i}-P)=0.
\label{eq:B16a} 
\end{equation}
Thus, every stable $SU(n)$ holomorphic vector bundle $V$ is
determined by specifying $n$ points $Q_{i}$ in $T^{2}$ up to permutations,
only $n-1$ of them being independent due to condition~\eqref{eq:B16a}.
Therefore, the moduli space of a stable $SU(n)$ holomorphic vector bundle
is
given by
\begin{equation}
{\cal M}_{bundle}(SU(n)) \simeq (T^2)^{n-1}/ \Sigma_n
\simeq {\mathbb C}{\mathbb P}^{n-1}.
\label{eq:B17}
\end{equation}
Comparing equations~\eqref{eq:B11} and~\eqref{eq:B17} we see that 
the two moduli spaces are identical. We conclude that, in a
straightforward
manner, the number of complex vector bundle moduli, $n-1$, and the moduli
space, ${\mathbb C}{\mathbb P}^{n-1}$, 
can be determined for a stable $SU(n)$ holomorphic
vector bundle on $T^{2}$. Furthermore, both the number of vector bundle
moduli
and their moduli space exactly correspond to the instanton moduli
determined
above by explicitly solving the Hermitian Yang-Mills equations.

Clearly the structure of moduli can be determined in two ways, either by
solving the Hermitian Yang-Mills equations or by constructing the
associated
holomorphic vector bundle and finding its deformations. In the simple case
of
the torus $T^{2}$, either method was efficacious. However, let us consider
complex manifolds $X$ where $dim_{{\mathbb C}}X \geq 2$. The Hermitian
Yang-Mills equations now become
\begin{equation}
F_{ab} = F_{\bar a \bar b} =0, \qquad g^{a \bar b}F_{a \bar b}=0.
\label{eq:B18}
\end{equation}
For flat manifolds $X$ of complex dimension two, the solutions of the
Hermitian
Yang-Mills equations can be constructed using the methods of ADHM 
\cite{ADHM}. However,
for non-flat twofolds and for all complex manifolds $X$ with 
$dim_{{\mathbb C}}X \geq 3$, there are no known solutions of these
equations.
Hence, the number and structure of moduli cannot be determined in this
way. However, the associated stable holomorphic vector bundles can be
constructed, at least for elliptically fibered Calabi-Yau manifolds using
the
methods of \cite{FMW1},\cite{AJ} and extended in 
\cite{gub}. Is it possible, therefore, that the
number and structure of moduli can be determined by studying the
deformations
of these holomorphic vector bundles? The answer is affirmative, as we will
show in detail in the remainder of this paper for stable $SU(n)$
holomorphic
bundles over elliptically fibered Calabi-Yau threefolds.


\section{Review of Spectral Covers and $SU(n)$ Vector \\ Bundles:}


\subsection*{Elliptically Fibered Calabi-Yau Threefolds:}

In this paper, we will consider Calabi--Yau threefolds, $X$, that are 
structured as elliptic curves fibered over a base surface,
$B$. We denote the natural projection as
$\pi: X\to B$  and by  $\sigma : B\to X$ the analytic map that defines the
zero
section.

A simple representation of an elliptic curve is given in the 
projective space $\cp{2}$ by the Weierstrass equation
\begin{equation}
zy^2=4x^3-g_2xz^2-g_3z^3,
\label{eq:1}
\end{equation}
where $(x,y,z)$ are the homogeneous coordinates of $\cp{2}$ and $g_2$,
$g_3$ are constants. This same equation can represent the elliptic 
fibration, $X$, if the coefficients $g_2$, $g_3$ in the Weierstrass
equation
are functions over the base surface, $B$. The correct way to express this 
globally is to replace the projective plane 
$\cp{2}$ by a $\cp{2}$-bundle $P \to B$ and then
require that $g_2$, $g_3$ be sections of appropriate line
bundles over the base. If we denote the conormal bundle to the 
zero section $\sigma(B)$ by $\cL$, then $P = {\mathbb P}({\mathcal
O}_{B}\oplus \cL^{2} \oplus \cL^{3})$, where ${\mathbb P}(W)$ stands
for the projectivization of a vector bundle $W$. There is a hyperplane
line
bundle ${\mathcal O}_{P}(1)$ on $P$ which corresponds to the divisor 
${\mathbb P}(\cL^{2}\oplus \cL^{3}) \subset P$ and the
coordinates $x,y,z$ are sections of
$\cO_{P}(1)\otimes\cL^2, \cO_{P}(1)\otimes
\cL^3$ and $\cO_{P}(1)$ respectively. 
It then follows from \eqref{eq:1} that the coefficients $g_2$ and $g_3$
are 
sections of $\cL^4$ and $ \cL^6$.

It is useful to define new coordinates, $\bX,\bY,\bZ$, on $X$ by
$x=\bX\bZ$,
$y=\bY$ and $z=\bZ^3$. It follows that $\bX,\bY,\bZ$ are now sections of
line bundles
\begin{equation}
\bX \sim \cO_{X}(2\sigma)\otimes\cL^2 , \qquad
\bY \sim \cO_{X}(3\sigma)\otimes \cL^3 , \qquad
\bZ \sim \cO_{X}(\sigma)
\label{eq:3}
\end{equation}
respectively, where we use the fact that $\cO_{X}(3\sigma)=\cO_{P}(1)$. 
The coefficients $g_2$ and $g_3$ remain 
sections of line bundles
\begin{equation}
g_2 \sim \cL^4, \qquad 
g_3 \sim \cL^6.
\label{eq:4}
\end{equation}
The symbol ``$\sim$'' simply means ``section of''.

The requirement that elliptically fibered threefold, $X$, be a
Calabi--Yau space 
constrains the first Chern class of the tangent bundle, $TX$, to
vanish. That is,
\begin{equation}
c_1(TX)=0.
\label{eq:5}
\end{equation}
It follows from this that
\begin{equation}
\cL=K_B^{-1},
\label{eq:6}
\end{equation}
where $K_B$ is the canonical bundle on the base, $B$. 
Condition~\eqref{eq:6} is
rather strong and restricts the allowed base spaces of an elliptically
fibered Calabi--Yau threefold to be del Pezzo, Hirzebruch and Enriques
surfaces, as well as certain blow--ups of Hirzebruch surfaces \cite{vafa}
.

\subsection*{Spectral Cover Description of $SU(n)$ Vector Bundles:}

As discussed in \cite{FMW1,AJ}, $SU(n)$ vector bundles over an
elliptically
fibered Calabi--Yau threefold can be explicitly constructed from two
mathematical objects, a divisor $\cC$ of $X$, called the spectral cover,
and a
line bundle $\cN$ on $\cC$. In this paper, we will 
consider only stable $SU(n)$ vector bundles
constructed from irreducible spectral covers. The moduli of 
semi--stable vector bundles associated with reducible spectral covers will 
be discussed elsewhere. In addition, we will impose the condition that the
spectral cover be a ``positive'' divisor.

\subsection*{Spectral Cover:}

A spectral cover, $\cC$, is a surface in $X$ that is an $n$-fold cover of
the
base $B$. That is, $\pi_{\cC}: \cC\to B$. The general form for a
spectral cover is given by
\begin{equation}
\cC=n\sigma + \pi^*\eta,
\label{eq:7}
\end{equation}
where $\sigma$ is the zero section and $\eta$ is some curve in the base
$B$.
The terms in~\eqref{eq:7} can be considered either as
elements of the homology group $H_{4}(X, {\mathbb Z})$ or, by
Poincare duality, as elements of cohomology $H^{2}(X, {\mathbb Z})$. 

In terms of the coordinates $\bX$, $\bY$, $\bZ$ introduced above, 
it can be shown that
the spectral cover can be represented as the zero set of the polynomial
\begin{equation}
s=a_{0}\bZ^{n} + a_{2}\bX\bZ^{n-2} + a_{3}\bY\bZ^{n-3} + \ldots +
a_{n}\bX^{\frac{n}{2}} 
\label{eq:8}
\end{equation}
for $n$ even and ending in $a_{n}\bX^{\frac{n-3}{2}}\bY$ if $n$ is odd,
along with the relation
s~\eqref{eq:3}. This tells us that the polynomial
$s$ must be a holomorphic section of the line bundle of the spectral 
cover,
$\cO_{X}(\cC)$. That is,
\begin{equation}
s \sim \cO_{X}(n\sigma + \pi^{*}\eta).
\label{eq:9}
\end{equation}
It follows from this and equations~\eqref{eq:3} and~\eqref{eq:4}, that the
coefficients $a_{i}$ in the polynomial $s$ must be sections of the line
bundles
\begin{equation}
a_{i} \sim \pi^*K_{B}^{i} \otimes \cO_{X}(\pi^{*}\eta)
\label{eq:10}
\end{equation}
for $i=1,\ldots ,n$ where we have used expression~\eqref{eq:6}.

In order to describe vector bundles most simply, there are two properties
that we require the spectral cover to possess. The first, which is shared
by
all spectral covers, is that 
\begin{itemize}
\item $\cC$ must be an effective class in $H_{4}(X,{\mathbb Z})$.
\end{itemize}
This property is simply an expression of the fact the spectral cover must
be
an actual surface in $X$. It can easily be shown that 
\begin{equation}
\cC \subset X \text{ is effective } \Longleftrightarrow \eta 
\text{ is an effective class in } H_{2}(B, {\mathbb Z}).
\label{eq:12}
\end{equation}
The second property that we require for the spectral cover is that
\begin{itemize}
\item $\cC$ is an irreducible surface.
\end{itemize}
This condition is imposed because it guarantees that the associated vector
bundle is stable. Deriving the 
conditions under which $\cC$ is irreducible is not completely trivial
and will be 
discussed elsewhere. Here, we will simply state the results.
First,  recall from~\eqref{eq:10} that 
$a_{i} \sim \pi^*K_{B}^{i}\otimes 
\cO_{X}(\pi^{*}\eta)$ and, hence, the zero locus of $a_{i}$ is a divisor,
$D(a_{i})$, in $X$. Then, we can show that $\cC$ is an irreducible surface
if
\begin{equation}
D(a_{0}) \text{ is an irreducible divisor in } X
\label{eq:13}
\end{equation}
and
\begin{equation}
D(a_{n}) \text{ is an effective class in } H_{4}(X, {\mathbb Z}).
\label{eq:14}
\end{equation}
Using Bertini's theorem, it can be shown that condition~\eqref{eq:13} is
satisfied if the linear system $|\eta|$ is base point free. 
``Base point free'' means that for any $b \in B$, we can find a
member of the linear system $|\eta|$  that does not pass through the point
$b$. 

In order to make these concepts more concrete, we take, as an example, the
base surface to be
\begin{equation}
B={\mathbb F}_{r}
\label{eq:15}
\end{equation}
and derive the conditions under  
which~\eqref{eq:12},~\eqref{eq:14} and~\eqref{eq:13}
are satisfied. Recall [] that the homology group $H_{2}({\mathbb
F}_{r}, {\mathbb Z})$ has
as a basis the effective classes $\cS$ and $\cE$ with intersection numbers
\begin{equation}
\cS^{2}=-r, \qquad \cS\cdot \cE=1, \qquad \cE^{2}=0.
\label{eq:16}
\end{equation}
Then, in general, $\cC$ is given by expression~\eqref{eq:7} where
\begin{equation}
\eta= a\cS+b\cE
\label{eq:17}
\end{equation}
and $a$,$b$ are integers. One can easily check that $\eta$ is an effective
class in ${\mathbb F}_{r}$, and, hence, that $\cC$ is an effective class
in
$X$, if and only if
\begin{equation}
a \geq 0, \qquad b \geq 0.
\label{eq:18}
\end{equation}
It is also not too hard to demonstrate that the linear system $|\eta|$ is
base
point free if and only if
\begin{equation}
b \geq ar.
\label{eq:19}
\end{equation}
Imposing this constraint then implies that $D(a_{0})$ is an irreducible
divisor in
$X$. Finally, we can show that for $D(a_{n})$ to be effective in $X$ one
must have
\begin{equation}
a \geq 2n, \qquad b \geq n(r+2).
\label{eq:20}
\end{equation}
Combining conditions~\eqref{eq:19} and~\eqref{eq:20} then guarantees that
$\cC$ is an irreducible surface. We now turn to the second
mathematical object that is required to specify an $SU(n)$ vector bundle.

In this paper, we require a third condition on the spectral cover that
\begin{itemize}
\item $\cC$ is positive.
\end{itemize}
This condition is imposed because, as we will see below, it allows one to
use
the Kodaira Vanishing theorem and the Lefschetz theorem when evaluating
the
number
of vector bundle moduli. These theorems greatly simplify this calculation.
By definition, $\cC$ is positive (or ample) if and only if the first Chern
class of the associated line bundle $\cO_{X}(\cC)$ is a positive class in
$H^{2}_{DR}(X, {\mathbb R})$. An equivalent, and from the point of view of
this paper more useful, definition of positive is the following. $\cC$ is
a
positive divisor if and only if
\begin{equation}
\cC \cdot c >0
\label{eq:20a}
\end{equation}
for every effective curve $c$ in $X$.

In order to make this concept more concrete, we take, as an example,
$B={\mathbb F}_{r}$ and $\eta=a\cS+b\cE$, where $a,b$ are integers. To
proceed, we must identify all of the effective curves $c$ in $X$. These
are
easily shown to be
\begin{equation}
(\pi^{*}\cS) \cdot \sigma, \qquad (\pi^{*}\cE) \cdot \sigma, \qquad F,
\label{eq:20b}
\end{equation}
where $F$ is the class of a generic fiber of $X$. First consider $\cC
\cdot
F$. Using
\begin{equation}
\sigma \cdot F=1, \qquad (\pi^{*}\cS) \cdot F = (\pi^{*}\cE) \cdot F=0, 
\label{eq:20c}
\end{equation}
it is straightforward to show that 
\begin{equation}
\cC \cdot F=n, 
\label{eq:20d}
\end{equation}
which is positive. The fact that $\sigma \cdot F=1$ is obvious. The
relation 
$(\pi^{*}\cS) \cdot F = (\pi^{*}\cE) \cdot F=0$ follows from the fact that
one
can always choose a representation of the class $F$ that does not
intersect
either $\pi^{*}\cS$ or $\pi^{*}\cE$. To evaluate $\cC \cdot ( (\pi^{*}\cS) 
\cdot \sigma)$ and $\cC \cdot ( (\pi^{*}\cE) \cdot \sigma)$ requires the
somewhat more subtle fact that
\begin{equation}
\sigma \cdot \sigma =-(\pi^{*}c_{1}(B)) \cdot \sigma,
\label{eq:20e}
\end{equation}
where $c_{1}(B)$ is the first Chern class of the base $B$. This can be
proven
as follows. First, note that by adjunction
\begin{equation}
K_{\sigma}= (K_{X} \otimes \cO_{X}(\sigma))|_{\sigma},
\label{eq:20f}
\end{equation}
where $K_{X}$ and $K_{\sigma}$ are the canonical line bundles on $X$ and
$\sigma \subset X$ respectively. Since $X$ is a Calabi-Yau threefold, it
follows from~\eqref{eq:5} that
\begin{equation}
c_{1}(K_{X}) =-c_{1}(TX)=0
\label{eq:20g}
\end{equation}
and, hence, that $K_{X}$ is the trivial bundle
\begin{equation}
K_{X} = \cO.
\label{eq:20h}
\end{equation}
Therefore, equation~\eqref{eq:20f} becomes
\begin{equation}
K_{\sigma} = \cO_{X}(\sigma)|_{\sigma}.
\label{eq:20i}
\end{equation}
Pulling this back onto the base $B$ using the map $\sigma:B \rightarrow X$
yields
\begin{equation}
K_{B} = \sigma^{*}\cO_{X}(\sigma)|_{\sigma},
\label{eq:20j}
\end{equation}
where $K_{B}$ is the canonical bundle of $B$. Therefore
\begin{equation}
c_{1}(K_{B}) = c_{1}(\sigma^{*}\cO_{X}(\sigma)|_{\sigma}).
\label{eq:20k}
\end{equation}
Using the relation
\begin{equation}
c_1 (K_B)  = - c_1 (B),
\label {eq:21}
\end{equation}
it follows that 
\begin{equation}
-c_1 (B)= c_1(\sigma^{*}\cO_{X}(\sigma)|_\sigma).
\label{eq:22}
\end{equation}
Pulling this expression back onto $\sigma$ using the map 
$\pi: X \rightarrow B$, one finds
\begin{equation}
-(\pi^* c_1(B)) \cdot \sigma =
(\pi^* c_1(\sigma^* \cO_{X} (\sigma)|_{\sigma} )) \cdot \sigma.
\label{eq:23}
\end{equation}
Note, however, that $c_1 (\sigma^{*} \cO_X (\sigma)|_{\sigma})$ is a curve
on $B$ constructed by pulling back the line bundle associated with
$\sigma$
restricted to itself. Therefore 
$(\pi^{*} c_1 (\sigma^{*} \cO_{X} (\sigma)|_{\sigma})) \cdot \sigma $
is precisely the curve $ \sigma \cdot \sigma $ in $X$, thus establishing
relation~\eqref{eq:20e}.

Using~\eqref{eq:20c},~\eqref{eq:20e} and the relation 
\begin{equation}
(\pi^{*} \gamma_1) \cdot (\pi^{*} \gamma_2)=\pi^{*} 
(\gamma_1 \cdot \gamma_2)
\label{eq:23a} 
\end{equation}
for any two curves $\gamma_1$, $\gamma_2$ on $B$, one can now evaluate
$\cC \cdot ((\pi^* \cS)\cdot \sigma)$ and
$\cC \cdot ((\pi^* \cE)\cdot \sigma)$. To do this, one should note that
for
\begin{equation}
B = {\mathbb F}_{r},
\label{eq:24}
\end{equation}
the Chern classes are known and given by
\begin{equation}
c_{1}({\mathbb F}_{r})=2\cS+(r+2)\cE, \qquad c_{2}({\mathbb F}_{r})=4.
\label{eq:25}
\end{equation}
Combining everything, we find that 
\begin{equation}
\cC \cdot ((\pi^* \cS)\cdot \sigma) = n (r-2)-ar +b,
\label{eq:25a}
\end {equation}
which will be positive if and only if 
\begin{equation}
b>ar-n(r-2).
\label{eq:26a}
\end{equation}
Similarly
\begin{equation}
\cC \cdot ((\pi^* \cE)\cdot \sigma) = -2n +a.
\label{eq:26b}
\end {equation}
For this to be positive one must have 
\begin{equation}
a>2n.
\label{eq:26c}
\end{equation}
We conclude that the spectral cover $\cC$ is positive if and only if 
conditions ~\eqref{eq:26a} and  ~\eqref{eq:26c} are satisfied. 

\subsection*{The Line Bundle $\cN$:}

As discussed in \cite{FMW1,AJ}, in addition to the 
spectral cover it is necessary to 
specify a line bundle, $\cN$, over $\cC$. For $SU(n)$ vector bundles, this
line bundle must be a restriction of a global line bundle on $X$
(which we will again denote by ${\mathcal N}$), satisfying the condition
\begin{equation}
c_{1}(\cN)=n(\frac{1}{2}+\lambda)\sigma+(\frac{1}{2}-\lambda)
\pi^{*}\eta+(\frac{1}{2}+n\lambda)\pi^{*}c_{1}(B),
\label{eq:27}
\end{equation}
where $c_{1}(\cN)$, $c_{1}(B)$ are the first Chern classes of $\cN$ and
$B$
respectively and $\lambda$ is, a priori, a rational number. Since
$c_{1}(\cN)$
must be an integer class, it follows that either
\begin{equation}
n \quad \mbox{is odd}, \qquad \lambda=m+\frac{1}{2}
\label{eq:28}
\end{equation}
or
\begin{equation}
n \quad \mbox{is even}, \qquad \lambda=m, \qquad \eta=c_{1}(B) \mod 2,
\label{eq:29}
\end{equation}
where $m \in {\mathbb Z}$. 

\subsection*{$SU(n)$ Vector Bundle:}

Given a spectral cover, $\cC$, and a line bundle, $\cN$, satisfying the
above
properties, one can now uniquely construct an $SU(n)$ vector bundle, $V$.
This
can be accomplished in two ways. First, as discussed in \cite{AJ}, the
vector
bundle
can be directly constructed using the associated Poincare bundle,
${\mathcal P}$.
The result is that 
\begin{equation}
V=\pi_{1*}(\pi^{*}_{2}\cN\otimes {\mathcal P}),
\label{eq:32}
\end{equation}
where $\pi_{1}$ and $\pi_{2}$ are the two projections of the fiber
product $X \times_{B} \cC$ onto the two factors $X$ and $\cC$. We refer
the reader to \cite{AJ,gub} for a detailed discussion. Equivalently, $V$
can be
constructed directly from $\cC$ and $\cN$ using the Fourier-Mukai
transformation, as discussed in \cite{FMW1,AJ}. Both of these
constructions work in
reverse, yielding the spectral data $(\cC, \cN)$ up to the overall factor
of $K_B$ given the vector bundle $V$. 
Throughout this paper we will indicate this relationship between the
spectral data and the vector bundle by writing
\begin{equation}
(\cC,\cN) \longleftrightarrow V.
\label{eq:33}
\end{equation}
The Chern classes for the $SU(n)$ vector bundle $V$ have been computed in
\cite{FMW1} and \cite{gub, smb}. The results are
\begin{equation}
c_{1}(V)=0
\label{eq:34}
\end{equation}
since $\operatorname{tr} F=0$ for the structure group $SU(n)$,
\begin{equation}
c_2(V)=\pi^{*}\eta\cdot\sigma-\frac{1}{24}\pi^{*}c_1(B)^2(n^3-n)
+\frac{1}{2}(\lambda^2-\frac{1}{4})n\pi^{*}(\eta\cdot(\eta-nc_1(B))) 
\label{eq:35}
\end{equation}
and
\begin{equation}
c_3(V)= 2\lambda \sigma\cdot \pi^{*}(\eta\cdot(\eta-nc_1(B))). 
\label{eq:36} 
\end{equation}
Finally, we note that it was shown in \cite{gub} that
\begin{equation}
N_{gen}=\frac{c_{3}(V)}{2}
\label{eq:36a}
\end{equation}
gives the number of quark and lepton generations.

To conclude, in this section we have discussed the construction and
properties
of stable $SU(n)$ vector bundles associated with positive, irreducible
spectral
covers over elliptically fibered Calabi-Yau threefolds.
For the remainder of this paper, for brevity, we will refer to such
bundles simply as ``stable $SU(n)$ vector bundles''. In order to make
these
concepts more transparent, we now present several examples.\\

\noindent $\bf Example \rm$ $\bf 1: \rm$ Consider a vector 
bundle specified by $B={\mathbb F}_{1}$,
$G=SU(3)$ and spectral cover
\begin{equation}
\cC=3\sigma+\pi^{*}\eta,
\label{eq:36b}
\end{equation}
where
\begin{equation}
\eta=7\cS+12\cE.
\label{eq:36c}
\end{equation}
That is, $r=1$, $n=3$, $a=7$, and $b=12$. Clearly, these parameters
satisfy~\eqref{eq:18} as well as~\eqref{eq:19},~\eqref{eq:20} and, hence,
$\cC$ is
effective and irreducible respectively. Furthermore,~\eqref{eq:26a}
and~\eqref{eq:26c} are satisfied, implying that $\cC$ is a positive
divisor. Now
choose the line bundle $\cN$ with
\begin{equation}
\lambda=\frac{1}{2},
\label{eq:36d}
\end{equation}
which satisfies condition~\eqref{eq:28} for $m=0$.
Using~\eqref{eq:16},~\eqref{eq:25},~\eqref{eq:36b} and~\eqref{eq:36c} we
find
from~\eqref{eq:36} and~\eqref{eq:36a} that
\begin{equation}
N_{gen}=13.
\label{eq:36e}
\end{equation}
is the number of generations.\\

\noindent $\bf Example \rm$ $\bf 2: \rm$ As a second example, 
we again choose $B={\mathbb F}_{1}$
and $G=SU(3)$, but now consider spectral cover 
\begin{equation}
\cC'=3\sigma+\pi^{*}\eta',
\label{eq:36bb}
\end{equation}
where
\begin{equation}
\eta'=7\cS+13\cE.
\label{eq:36cc}
\end{equation}
That is, $r=1$, $n=3$, $a'=7$, and $b'=13$, which clearly 
satisfy~\eqref{eq:18},~\eqref{eq:19},~\eqref{eq:20} 
and~\eqref{eq:26a},~\eqref{eq:26c}. Hence $\cC'$ is an effective,
irreducible,positive divisor. We again take
\begin{equation}
\lambda=\frac{1}{2}.
\label{eq:36dd}
\end{equation}
Using~\eqref{eq:16},~\eqref{eq:25},~\eqref{eq:36bb} and~\eqref{eq:36cc} we
find
from~\eqref{eq:36} and~\eqref{eq:36a} that
\begin{equation}
N_{gen}=17.
\label{eq:36ee}
\end{equation}
is the number of generations.\\

\noindent $\bf Example \rm$ $\bf 3: \rm$ As a final example, 
we again take $B={\mathbb F}_{1}$
and $G=SU(3)$, but consider 
\begin{equation}
\cC'=3\sigma+\pi^{*}\eta',
\label{eq:36bbb}
\end{equation}
where
\begin{equation}
\eta'=8\cS+12\cE.
\label{eq:36ccc}
\end{equation}
That is, $r=1$, $n=3$, $a'=8$, and $b'=12$, which clearly 
satisfy~\eqref{eq:18},~\eqref{eq:19},~\eqref{eq:20} 
and~\eqref{eq:26a},~\eqref{eq:26c}. Hence $\cC'$ is an effective,
irreducible,positive divisor. Again, let
\begin{equation}
\lambda=\frac{1}{2}.
\label{eq:36ddd}
\end{equation}
Using~\eqref{eq:16},~\eqref{eq:25},~\eqref{eq:36bbb} and~\eqref{eq:36ccc}
we find
from~\eqref{eq:36} and~\eqref{eq:36a} that
\begin{equation}
N_{gen}=16.
\label{eq:36eee}
\end{equation}
is the number of generations.\\

We will refer to these examples elsewhere in this paper to illustrate our
results on vector bundle moduli.


\section{The Moduli of Stable $SU(n)$ Vector Bundles:}


The moduli of holomorphic vector bundles have been discussed, in a generic
way, in \cite{FMW1,AJ} and elsewhere. For example, general properties of
heterotic vector bundle moduli on certain elliptically fibered Calabi-Yau
spaces, as well as on the dual $F$-theory compactifications, have been
discussed in \cite{1,2,3,4,5,6,7}. In this section, we present an explicit
calculation of the number of moduli of stable, irreducible
$SU(n)$ holomorphic vector bundles over elliptically fibered Calabi-Yau
threefolds. For specificity, we present an explicit formula for the number of
vector bundle moduli when the Calabi-Yau threefold has base $B={\mathbb F}_{r}$.

It is clear from expression~\eqref{eq:33} that the number of moduli of a
stable $SU(n)$ vector bundle $V$ is determined by the number of parameters
specifying its spectral cover $\cC$ and by the size of the space of
holomorphic
line bundles $\cN$ defined on $\cC$. We begin by 
considering the spectral cover.

\subsection*{Parameters of $\cC$:}

The spectral cover $\cC$ is a divisor of the Calabi-Yau threefold $X$ and,
hence, uniquely determines a line bundle $\cO_{X}(C)$ on $X$. Then $H^{0}
(X,\cO_{X}(C))$ is the space of holomorphic sections of $\cO_{X}(C)$. We
denote
its dimension by $h^{0}(X,\cO_{X}(C))$. It follows that there must exist
$h^{0}$ holomorphic sections $s_{1},..,s_{h^{0}}$ that span this space.
Note
that the zero locus of each holomorphic section of the form 
\begin{equation}
 s_{\{a_{i}\}}=\Sigma_{i=1}^{h^{0}}a_{i}s_{i},
\label{eq:37}
\end{equation}
for fixed complex coefficients $a_{i}$, determines an effective divisor
$\cC_{\{a_{i}\}}$ of $X$. Running over all $a_{i}$ gives the set,
$|\cC|$, of effective divisors associated with $\cO_{X}(C)$. Clearly
\begin{equation}
 |\cC|={\mathbb P}H^{0}(X,\cO_{X}(C)),
\label{eq:38}
\end{equation}
where the right hand side is the projectivization of
$H^{0}(X,\cO_{X}(C))$. It
follows that 
\begin{equation}
 dim|\cC|=h^{0}(X,\cO_{X}(\cC))-1.
\label{eq:39}
\end{equation}
This quantity counts the number of parameters specifying the
spectral
cover $\cC$. We now consider the line bundles $\cN$ over $\cC$.

\subsection*{The Space of Line Bundles $\cN$:}

The set of holomorphic line bundles $\cN$ over the spectral cover $\cC$ 
is, by definition, determined by the set of holomorphic transition
functions allowed on $\cC$. These, in turn, are specified as the closed
but not
exact elements of the multiplicative group $C^{1}(\cC,\cO_{\cC}^{*})$ of 
non-vanishing holomorphic functions on the intersection of any two open
sets in
the atlas of $\cC$. That is, the group of line bundles of $\cC$ is given
by
\begin{equation}
 Pic(\cC)=H^{1}(\cC,\cO_{\cC}^{*}),
\label{eq:40}
\end{equation}
where $ H^{1}(\cC,\cO_{\cC}^{*})$ is the first \v Cech cohomology group of
$ \cO_{\cC}^{*}$ on $\cC$. Clearly then, the size of the space of line
bundles $\cN$ over $\cC$ is specified by
\begin{equation}
 dimPic(\cC)=h^{1}(\cC,\cO_{\cC}^{*}).
\label{eq:41}
\end{equation}
\\
Putting the results of the last two subsections together, we see that the
number of moduli of a stable $SU(n)$ vector bundle $V$, which we will
denote
by $n(V)$, is given by
\begin{equation}
 n(V)= (h^{0}(X,\cO_{X}(\cC))-1) + h^{1}(\cC,\cO_{\cC}^{*}).
\label{eq:42}
\end{equation}
We now turn to the explicit evaluation of each of the terms in this
expression.

\subsection*{The Evaluation of $h^{0}(X,\cO_{X}(\cC))$:}

Our basic approach to evaluating $h^{0}(X,\cO_{X}(\cC))$ is through the
Riemann-Roch theorem which, in this context, states that
\begin{equation}
\chi_{E}(X,\cO_{X}(\cC))=h^{0}(X,\cO_{X}(\cC))-h^{1}+h^{2}-h^{3},
\label{eq:43}
\end{equation}
where $\chi_{E}(X,\cO_{X}(\cC))$ is the Euler characteristic and $h^{q}$
for
$q=1,2,3$ are the dimensions of the higher cohomology groups of $X$
evaluated in $\cO_{X}(\cC)$. To proceed, we will use the fact that the 
stable $SU(n)$ vector bundles that we consider in this
paper are positive. This allows us to employ the Kodaira vanishing theorem
which states that, for positive line bundle $\cO_{X}(\cC)$ and
$dim_{{\mathbb
C}}X=m$,
\begin{equation}
H^{q}(X,\Omega_{X}^{p}(\cO_{X}(\cC))=0
\label{eq:44}
\end{equation}
for $p+q>m$. In our case $m=3$ and consider $p=3$. Then
\begin{equation}
\Omega_{X}^{3}(\cO_{X}(\cC))=\Omega_{X}^{3}(TX) \otimes \cO_{X}(\cC).
\label{eq:45}
\end{equation}
However, since we are on a threefold
\begin{equation}
\Omega_{X}^{3}(TX)=K_{X},
\label{eq:46}
\end{equation}
where $K_{X}$ is the canonical bundle of $X$. But $X$ is a Calabi-Yau
manifold and, hence, $K_{X}$ is the trivial line bundle.Therefore,
\begin{equation}
\Omega_{X}^{3}(\cO_{X}(\cC))=\cO_{X}(\cC).
\label{eq:47}
\end{equation}
In this situation, the vanishing theorem then states that
\begin{equation}
H^{q}(X,\cO_{X}(\cC))=0
\label{eq:48}
\end{equation}
for $q>0$. It follows that $h^{q}=0$ for $q=1,2,3$ and the Riemann-Roch
theorem simplifies to 
\begin{equation}
\chi_{E}(X,\cO_{X}(\cC))=h^{0}(X,\cO_{X}(\cC)).
\label{eq:49}
\end{equation}
Therefore, to evaluate $h^{0}(X,\cO_{X}(\cC))$ we need simply to evaluate
the
Euler characteristic. For the situation at hand, the Euler characteristic
is determined from the Atiyah-Singer index theorem to be
\begin{equation}
\chi_{E}(X,\cO_{X}(\cC))=\int_{X}{ch(\cO_{X}(\cC)) \wedge Td(TX)},
\label{eq:50}
\end{equation}
where $ch$ and $Td$ are the total Chern character and Todd class
respectively.
This can be evaluated for any elliptically fibered Calabi-Yau threefold of
base
$B$. However, throughout this paper, we will do explicit calculations only 
for the case where 
\begin{equation}
B={\mathbb F}_{r}.
\label{eq:51}
\end{equation}
Recall from the previous section that, in this case, the spectral cover
has the generic form
\begin{equation}
\cC=n\sigma + \pi^{*}(a\cS + b\cE),
\label{eq:52}
\end{equation}
where $a,b$ are integers. For the present calculation, we need put no 
restrictions on $a$ and $b$. On a Calabi-Yau threefold,~\eqref{eq:50}
takes a form 
\begin{equation}
\label{eq:50a}
\chi_{E}(X,\cO_{X}(\cC))= 
\frac{1}{6} \int_{X}c_{1}^{3}(\cO_{X}(\cC))+
\frac{1}{12}\int_{X} c_{1}(\cO_{X}(\cC)) \wedge c_2(TX).
\end{equation}
The first Chern class $c_1(\cO_{X} (\cC))$ is just given by $\cC$.
The second Chern class of the tangent bundle $c_2(TX)$ has been found in 
\cite{FMW1} to be
\begin{equation}
c_2(TX)= \pi^{*}(c_2(B)+ 11 c_1^2 (B))+12 \sigma \cdot \pi^{*}(c_1 (B)).
\label{eq:50b}
\end{equation}
Now, by using equation~\eqref{eq:25} for the first and second Chern
classes of ${\mathbb F}_{r}$ and equation~\eqref{eq:20e} one finds
\begin{equation}
\chi_{E}(X,\cO_{X}(\cC))=\frac{n}{3}(4n^{2}-1)+nab-(n^{2}-2)(a+b)
+ar(\frac{n^{2}}{2}-1)-\frac{n}{2}ra^{2}.
\label{eq:53}
\end{equation}
Of course, there are further restrictions on the integers $a$ and $b$.
These
are 1) the non-negativity conditions~\eqref{eq:18} required to make $\cC$
effective and 2) the conditions~\eqref{eq:19},~\eqref{eq:20}
and~\eqref{eq:26a},~\eqref{eq:26b} 
necessary to render $\cC$ irreducible and positive. 
Under these conditions, equation~\eqref{eq:49} is valid and, hence,
\begin{equation}
h^{0}(X,\cO_{X}(\cC))=\frac{n}{3}(4n^{2}-1)+nab-(n^{2}-2)(a+b)
+ar(\frac{n^{2}}{2}-1)-\frac{n}{2}ra^{2}.
\label{eq:54}
\end{equation}
We now proceed to calculate $h^{1}(\cC,\cO_{\cC}^{*})$.

\subsection*{The Evaluation of $h^{1}(\cC,\cO_{\cC}^{*})$:} 

In order to evaluate $h^{1}(\cC,\cO_{\cC}^{*})$ it will first be necessary
to
construct some related cohomology groups. The Hurewicz theorem tells us
that
\begin{equation}
H_{1}(X, {\mathbb Z})=\pi_{1}(X)/[\pi_{1}(X),\pi_{1}(X)],
\label{eq:55}
\end{equation}
where $[\pi_{1}(X),\pi_{1}(X)]$ is the commutator subgroup of
$\pi_{1}(X)$.
For most of the Calabi-Yau threefolds of interest to particle physics, the
fundamental group will be finite, such as  $\pi_{1}={\bf 1}$,
${\mathbb Z}_{2}$, ${\mathbb Z}_{2} \times {\mathbb Z}_{2}$ and so on. It
follows that
\begin{equation}
H^{1}_{DR}(X,{\mathbb R})=0.
\label{eq:56}
\end{equation}
However,
\begin{equation}
H^{1}_{DR}(X,{\mathbb R}) \otimes {\mathbb C} \equiv H^{1}_{DR}(X,{\mathbb
C}) =H^{1,0}_{\bar{\partial}}(X,{\mathbb C}) \oplus
H^{0,1}_{\bar{\partial}}
(X,{\mathbb C}).
\label{eq:57}
\end{equation}
We see from this and~\eqref{eq:56} that
\begin{equation}
H^{0,1}_{\bar{\partial}}(X,{\mathbb C})=0.
\label{eq:58}
\end{equation}
Finally, using the Dolbeault theorem 
\begin{equation}
H^{0,1}_{\bar{\partial}}(X,{\mathbb C}) \cong H^{1}(X,\cO_{X}),
\label{eq:59}
\end{equation}
we conclude that
\begin{equation}
H^{1}(X,\cO_{X})=0.
\label{eq:60}
\end{equation}
However, the group $H^{1}(X,\cO_{X})$ is not quite what we want. What we
really need to know is the restriction of this group to the spectral cover
$\cC$. To find this restriction, we can use the Lefschetz theorem which
states the
following. For a compact $m$-fold $X$ and a smooth,
positive $m-1$ dimensional submanifold $\cC \subset X$, then the map
\begin{equation}
H^{p}(X,\Omega_{X}^{q}) \rightarrow H^{p}(\cC,\Omega_{\cC}^{q})
\label{eq:61}
\end{equation}
induced by the inclusion $i:\cC \rightarrow X$
is an isomorphism for $p+q \leq m-2$ and is injective for $p+q=m-1$.
Choosing
$p=1$, $q=0$ satisfies the isomorphism condition for $m=3$. Then, using
the fact that $\Omega^{0} \cong \cO$, the Lefschetz theorem becomes
\begin{equation}
H^{1}(X,\cO_{X}) \cong H^{1}(\cC,\cO_{\cC}).
\label{eq:62}
\end{equation}
It follows from this and~\eqref{eq:60} that
\begin{equation}
H^{1}(\cC,\cO_{\cC})=0.
\label{eq:63}
\end{equation}
This is still not quite what we require since we need the first cohomology
group evaluated over not $\cO_{\cC}$ but, rather, $\cO_{\cC}^{*}$. This
can
be obtained as follows. Recall that there is a simple exact sequence
\begin{equation}
0 \rightarrow {\mathbb Z} \rightarrow \cO_{\cC} \rightarrow \cO_{\cC}^{*}
\rightarrow 0,
\label{eq:64}
\end{equation}
where the map from ${\mathbb Z} \rightarrow \cO_{\cC}$ is inclusion and 
$\cO_{\cC} \rightarrow \cO_{\cC}^{*}$ is the exponential mapping defined
by $exp(f)=e^{2 \pi if}$. In the usual way, this produces a long exact
sequence of cohomology groups given in part by
\begin{equation}
\rightarrow H^{1}(\cC,\cO_{\cC}) \rightarrow H^{1}(\cC, \cO_{\cC}^{*})
\rightarrow H^{2}(\cC,{\mathbb Z}) \rightarrow,
\label{eq:65}
\end{equation}
where $H^{1}(\cC, \cO_{\cC}^{*}) \rightarrow H^{2}(\cC,{\mathbb Z})$ is
the \v Cech coboundary operation $\delta_{1}$. But, by
equation~\eqref{eq:63}
$H^{1}(\cC,\cO_{\cC})$ vanishes. Hence, the mapping
\begin{equation}
\delta_{1}: H^{1}(\cC, \cO_{\cC}^{*}) \rightarrow H^{2}(\cC,{\mathbb Z})
\label{eq:66}
\end{equation}
is injective. That is,
\begin{equation}
Pic(\cC)=H^{1}(\cC, \cO_{\cC}^{*}) \subset H^{2}(\cC,{\mathbb Z}).
\label{eq:67}
\end{equation}
Note that $H^{2}(\cC,{\mathbb Z})$ forms a rigid lattice and, hence, there
are no smooth deformations, as there would be in a space over the complex
numbers
${\mathbb C}$. We conclude that
\begin{equation}
dimH^{1}(\cC, \cO_{\cC}^{*})=0.
\label{eq:68}
\end{equation}

\subsection*{n(V) for $B={\mathbb F}_{r}$:}

We can now give the final expression for the number of moduli 
of a positive, stable $SU(n)$
vector bundle over an elliptically fibered Calabi-Yau threefold with base
$B={\mathbb F}_{r}$. The associated spectral cover is given by
\begin{equation}
\cC=n\sigma+\pi^{*}(a\cS +b\cE),
\label{eq:69}
\end{equation}
where the effectiveness of $\cC$ requires
\begin{equation}
a \geq 0, \qquad b \geq 0,
\label{eq:70}
\end{equation}
the irreducibility of $\cC$ demands that
\begin{equation}
b \geq ar, \qquad a \geq 2n, \qquad b \geq n(r+2)
\label{eq:71}
\end{equation}
and the positivity requirement for $\cC$ implies
\begin{equation}
a > 2n, \qquad b > ar-n(r-2).
\label{eq:72}
\end{equation}
Using equations~\eqref{eq:42},~\eqref{eq:54} and~\eqref{eq:68}, one can
conclude the following.

\begin{itemize} 

\item The number of vector bundle moduli is given by
\begin{equation}
n(V)=\frac{n}{3}(4n^{2}-1)+nab-(n^{2}-2)(a+b)
+ar(\frac{n^{2}}{2}-1)-\frac{n}{2}ra^{2}-1.
\label{eq:73}
\end{equation}
\end{itemize}

This equation will be 
essential to the subsequent discussion. To make this result more concrete,
we
evaluate it for each of the three stable $SU(3)$ vector bundles presented
at
the end of Section 3.\\

\noindent $\bf Example \rm$ $\bf 1: \rm$ In this case $B={\mathbb F}_{1}$,
$G=SU(3)$, $a=7$ and $b=12$. It follows
from~\eqref{eq:73} that the number of vector bundle moduli of $V$ is
\begin{equation}
n(V)=104.
\label{eq:73a}
\end{equation}
\\

\noindent $\bf Example \rm$ $\bf 2: \rm$ In this case $B={\mathbb F}_{1}$,
$G=SU(3)$, 
$a'=7$ and $b'=13$. 
Then~\eqref{eq:73} implies that
\begin{equation}
n(V')=118
\label{eq:73b}
\end{equation}
is the number of vector bundle moduli of $V'$.\\

\noindent $\bf Example \rm$ $\bf 3: \rm$ In this case $B={\mathbb F}_{1}$, 
$G=SU(3)$, $a'=8$ and $b'=12$. Then, it follows from~\eqref{eq:73} that
\begin{equation}
n(V')=114
\label{eq:73c}
\end{equation}
is the number of vector bundle moduli of $V'$.


\section{Moduli and Small Instanton Transitions:}


Heterotic $M$-theory \cite{ssea,losw1} is the compactification 
of Ho\v rava-Witten theory \cite{hw1}
on a smooth Calabi-Yau threefold with non-vanishing $G$-flux. At energies
below the scale of the Calabi-Yau space, heterotic $M$-theory appears as a
five-dimensional bulk space bounded at either end of the fifth dimension
by a four-dimensional end-of-the-world 3-brane. Specifically, 
these bounding branes are  
ten-dimensional $S^{1}/{\mathbb Z}_{2}$ orbifold fixed planes with six
spacelike
dimensions wrapped on the Calabi-Yau threefold. As shown by Ho\v rava and
Witten \cite{hw1}, anomaly cancellation demands that there must exist an
$E_{8}$,
$N=1$ supergauge multiplet on each orbifold plane. Compactification 
allows for the possibility that there is a non-trivial
vacuum $E_{8}$ gauge configuration on the Calabi-Yau manifold at each
end-of-the-world orbifold plane. If this vacuum ``instanton'' has
structure
group $G \subset E_{8}$, then the theory that appears on the low energy
3-brane worldvolume is an $N=1$ gauge theory with gauge group $H$, where
$H$
is
the commutant of $G$ in $E_{8}$. In addition, there appear matter
supermultiplets with a calculable number of families. Such $G$-instantons,
within the context of elliptically fibered and torus fibered Calabi-Yau
threefolds, are discussed in detail in \cite{gub,gub2,smb}. 

The vector bundles asociated with
the two end-of-the-world instantons are not independent. They are
non-trivially correlated by the requirement that the entire theory be
anomaly
free. As discussed in \cite{gub}, generally, anomaly freedom is difficult
to
satisfy
for phenomenologically interesting heterotic $M$-theories 
with end-of-the-world
branes alone. However, anomaly freedom is easily achieved if one allows
for
$M$5-branes in the bulk space \cite{hw1,gub,gub2}. 
These branes must be wrapped on an
effective
holomorphic curve in the Calabi-Yau manifold $X$ to give $N=1$
supersymmetric
bulk 3-branes. If we denote the Poincare dual of this holomorphic curve by
the
four-form $W$, then anomaly cancellation tells us that
\begin{equation}
W=c_{2}(TX)-c_{2}(V_{1})-c_{2}(V_{2}),
\label{eq:74}
\end{equation}
where $c_{2}(TX)$ is the second Chern class of the tangent bundle of $X$
and $c_{2}(V_{1})$, $c_{2}(V_{2})$ are the second Chern classes of the
first
and second end-of-the-world vector bundles respectively. Restricting $X$
to be
elliptically fibered, as we will henceforth do, $W$ decomposes as
\begin{equation}
W= W_{B} \cdot \sigma+a_{f}F,
\label{eq:75}
\end{equation}
where $W_{B}=\pi^{*}w$ is the lift of a curve $w$ in the base $B$ and
$a_{f}$
is an integer. That is, the holomorphic curve $W$ decomposes into a purely
horizontal curve $W_{B} \cdot \sigma$ and a purely vertical curve
proportional to the fiber $F$. It was shown in \cite{gub} that $W$ is an
effective
class in $H_{2}(X,{\mathbb Z})$ if and only if $w$ is an effective class
in
$H(B, {\mathbb Z})$ and $a_{f} \geq 0$, for any del Pezzo or Enriques base
$B$. This is also true for any 
Hirzebruch base $B={\mathbb F}_{r}$, except when
$w$ contains the negative section $\cS$ and $r \geq 3$. We will avoid this
pathological case in this paper.

Let us now assume that we have a vacuum configuration of heterotic
$M$-theory
consisting of two end-of-the-world branes with vector bundles $V_{1}$ and
$V_{2}$ respectively and internal bulk $M$5-branes wrapped on holomorphic
curve $W=(\pi^{*}w) \cdot \sigma + a_{f}F$. Let the spectral cover
describing
the $SU(n)$ vector bundle $V_{1}$ be given by
\begin{equation}
\cC=n\sigma+\pi^{*}\eta,
\label{eq:76}
\end{equation}
where we will always take $\cC$ to be effective, irreducible and positive.
Let $z$ be any effective, irreducible curve which is a component of some
representative of the class $w$. 
Then, it was shown in \cite{tony} that one can move the
curve $z$ to a boundary brane, we assume it is the first brane, and
``absorb''
it via a small instanton transition into the vector bundle $V_{1}$. This
transition results in
a new vector bundle $V_{1}'$. Henceforth, we will denote these two bundles
by
$V$ and $V'$ repectively. This process of absorbing all, or part, of the
horizontal component of the bulk $M$5-brane curve, passing first through a
torsion free sheaf (the small instanton) and then smoothing out to a
vector
bundle, was described in detail in \cite{tony}. As shown 
in \cite{tony}, the effect of this
transition, in addition to removing the $z$ component of the bulk
$M$5-brane
curve, is to modify the spectral cover~\eqref{eq:76} describing $V$ to a
new
spectral cover
\begin{equation}
\cC'=n\sigma+\pi^{*}(\eta+z)
\label{eq:77}
\end{equation}
which describes vector bundle $V'$. This transition preserves the low
energy
gauge group $H$, but changes the third 
Chern class of the vector bundle and, hence,
the number of families on the first 3-brane. Some of the physics of these
chirality changing small instanton transitions are discussed in
\cite{tony}.
Finally,
we note that it was demonstrated in \cite{tony} that 
this type of phase transition does not alter the line bundle
$\cN$ on the Calabi-Yau threefold.

Before continuing, we emphasize that it is
also possible to have a small instanton transition in which a purely
vertical
component of the $M$5-brane curve is absorbed \cite{tony}. However, at
least one of 
the bundles involved in such transitions is reducible and, hence, our
previous
discussion of moduli does not apply. We reserve discussion of such
transitions for a
later publication.

\subsection*{Transition Moduli:}

In the previous section, we showed how to compute the number of moduli,
$n(V)$, of any positive, stable holomorphic $SU(n)$ vector bundle over an
arbitrary
elliptically fibered Calabi-Yau threefold. 
For specificity, we computed $n(V)$ for 
$B={\mathbb F}_{r}$. In this section, we are interested in the 
``new'' moduli that appear when the vector bundle $V$ on the first brane 
is struck by an $M$5-brane wrapped on a horizontal curve and makes a small
instanton transition to bundle $V'$. We will refer to these as
``transition'' moduli. There are two issues to be discussed in
this regard. The first is to compute the number, $n_{tm}$, of these 
transition moduli. The second is to elucidate the exact mathematical and
physical relationship of the transition moduli to the torsion free sheaf and
the horizontal curve. We begin by computing the number of transition moduli.
This is clearly given by
\begin{equation}
n_{tm}=n(V')-n(V).
\label{eq:78}
\end{equation}
Although this number can be computed on a Calabi-Yau threefold with any
allowed base $B$, in this paper we will present the calculation for 
$B={\mathbb F}_{r}$. The result follows immediately from
equations~\eqref{eq:73} and~\eqref{eq:77}. If we take
\begin{equation}
\eta=a\cS+b\cE, \qquad \eta+z=a'\cS+b'\cE,
\label{eq:79}
\end{equation}
where both associated spectral covers $\cC$ and $\cC'$ are 
assumed to be effective,
irreducible and positive, then we can conclude the following.

\begin{itemize}

\item The number of transition moduli is given by
\begin{equation}
n_{tm}=n(a'b'-ab)-(n^{2}-2)(a'-a+b'-b)+(a'-a)r(\frac{n^{2}}{2}-1)
-\frac{nr}{2}((a')^{2}-a^{2}).
\label{eq:80}
\end{equation}

\end{itemize}

\noindent Generically, this is a non-vanishing positive integer indicating
that, at
the time of the small instanton transition, new moduli are added to the
holomorphic vector bundle. It is useful to illustrate these concepts by
presenting several concrete examples.\\

\noindent $\bf Example \rm$ $\bf 1: \rm$ Consider the 
stable $SU(3)$ vector bundle $V$ specified
in Example 1 at the end of Section 3. In this case, $B={\mathbb F}_{1}$,
$G=SU(3)$, $a=7$, $b=12$ and $\lambda=1/2$. As shown in~\eqref{eq:36e} 
and~\eqref{eq:73a}, $N_{gen}=13$ and $n(V)=104$ respectively. Using 
expressions~\eqref{eq:16},~\eqref{eq:25},~\eqref{eq:36},~\eqref{eq:50b}
and
assuming that the vector bundle $V_{2}$ over the second end-of-the-world
brane
is trivial, we find from ~\eqref{eq:74} and ~\eqref{eq:75} that the bulk
space
$M$5-brane is non-vanishing and wrapped on a holomorphic curve $W \subset
X$
where
\begin{equation}
W_{B}=\pi^{*}(17\cS+20\cE), \qquad a_{f}=100.
\label{eq:80a}
\end{equation}
Note that $W$ is an effective class in $H_{2}(X, {\mathbb Z})$, as it must
be.
Using the results of \cite{gub2}, one can show that there is always a
region of the
moduli space of the $M$5-brane where
\begin{equation}
z=\cE \subset 17\cS+20\cE
\label{eq:80b}
\end{equation}
is an effective subcurve. Curve $z=\cE$ can move to the first
end-of-the-world
brane, inducing a small instanton transition to a new vector bundle $V'$
specified by $B={\mathbb F}_{1}$, $G=SU(3)$, $a'=7$, $b'=13$ and
$\lambda=1/2$, where we have used~\eqref{eq:77}. But, this is precisely
the
stable $SU(3)$ vector bundle $V'$ presented in Example 2 at the end of
Section
3. As shown in~\eqref{eq:36ee} and~\eqref{eq:73b}, for this bundle
$N_{gen}'=17$ and $n(V')=118$. Using~\eqref{eq:80}, we find that the
number of
transition moduli is
\begin{equation}
n_{tm}=14.
\label{eq:80c}
\end{equation}
That is, the small instanton transition with $M$5-brane curve $z=\cE$
increases the number of vector bundle moduli from $n(V)=104$ before the
collision to $n(V')=118$ after it. Note, in passing, that this transition
also
changes the number of generations from $N_{gen}=13$ to $N_{gen}'=17$.\\

\noindent $\bf Example \rm$ $\bf 2: \rm$ Consider again the 
stable $SU(3)$ vector bundle $V$ specified
in Example 1 at the end of Section 3, with $B={\mathbb F}_{1}$,
$G=SU(3)$, $a=7$, $b=12$, $\lambda=1/2$ and $N_{gen}=13$, $n(V)=104$.
Assuming that the vector bundle $V_{2}$ over the second 
end-of-the-world brane is trivial, the bulk space
$M$5-brane is non-vanishing and wrapped on the holomorphic curve $W$
specified
in~\eqref{eq:80a}. Using the results of \cite{gub2}, 
one can show that there is always
another region of the moduli space of the $M$5-brane where
\begin{equation}
z=\cS \subset 17\cS+20\cE
\label{eq:80bb}
\end{equation}
is an effective subcurve. Curve $z=\cS$ can move to the first
end-of-the-world
brane, inducing a small instanton transition to a new vector bundle $V'$
specified by $B={\mathbb F}_{1}$, $G=SU(3)$, $a'=8$, $b'=12$ and
$\lambda=1/2$, where we have used~\eqref{eq:77}. But, this is precisely
the
stable $SU(3)$ vector bundle $V'$ presented in Example 3 at the end of
Section
3. As shown in~\eqref{eq:36eee} and~\eqref{eq:73c}, for this bundle
$N_{gen}'=16$ and $n(V')=114$. Using~\eqref{eq:80}, we find that the
number of
transition moduli is
\begin{equation}
n_{tm}=10.
\label{eq:80cc}
\end{equation}
That is, the small instanton transition with $M$5-brane curve $z=\cS$
increases the number of vector bundle moduli from $n(V)=104$ before the
collision to $n(V')=114$ after it. Note, in passing, that this transition
also
changes the number of generations from $N_{gen}=13$ to $N_{gen}'=16$.\\

We now turn to the second important question
regarding transion moduli. That is, what is the exact 
relationship of the transition moduli to the torsion free sheaf and the
horizontal curve?

\subsection*{Localization of the Transition Moduli on $\pi^{*}z$:}

One could well speculate that the transition moduli arise as the moduli of
the
spectral cover $\cC'$ restricted to the lift of the horizontal curve,
$\pi^{*}z$. We claim that this is, in fact, correct, as we now demonstrate
for
the case of $B={\mathbb F}_{r}$. 
For a divisor $D$ in $X$, there is a short exact
sequence
\begin{equation}
0 \rightarrow E \otimes \cO_{X}(-D) \rightarrow E \rightarrow
E|_{D} \rightarrow 0,
\label{eq:81}
\end{equation}
where $E$ is any holomorphic vector bundle on $X$. Taking $E$ to be
trivial,
$E=\cO$, implies that
\begin{equation}
0 \rightarrow \cO_{X}(-D) \rightarrow \cO_{X} \rightarrow
\cO_{D} \rightarrow 0.
\label{eq:82}
\end{equation}
We begin by choosing
\begin{equation}
z=\cE
\label{eq:83}
\end{equation}
and
\begin{equation}
D=\pi^{*}\cE,
\label{eq:84}
\end{equation}
which is a divisor in $X$. Then exact sequence~\eqref{eq:82} becomes
\begin{equation}
0 \rightarrow \cO_{X}(-\pi^{*}\cE) \rightarrow \cO_{X} \rightarrow
\cO_{\pi^{*}\cE} \rightarrow 0.
\label{eq:85}
\end{equation}
Tensoring each term with the line bundle
$\cO_{X}(n\sigma+\pi^{*}(a\cS+(b+1)\cE))$ yields
\begin{equation}
0 \rightarrow L \rightarrow L' \rightarrow
L'|_{\pi^{*}\cE} \rightarrow 0,
\label{eq:86}
\end{equation}
where
\begin{equation}
L=\cO_{X}(\cC), \qquad L'=\cO_{X}(\cC')
\label{eq:87}
\end{equation}
and
\begin{equation}
\cC=n\sigma+\pi^{*}(a\cS+b\cE), \qquad
\cC'=n\sigma+\pi^{*}(a\cS+(b+1)\cE).
\label{eq:87a}
\end{equation}
Note that $L'|_{\pi^{*}\cE}$ is the
line
bundle associated with the restriction of $\cC'$ to $\pi^{*}z$ and, hence,
is the object of interest in the conjecture. 

In passing, note that
\begin{equation}
L'|_{\pi^{*}\cE}=L|_{\pi^{*}\cE}.
\label{eq:88}
\end{equation}
To see this, use the fact 
\begin{equation}
L'=L \otimes \cO_{X}(\pi^{*}\cE)
\label{eq:89}
\end{equation}
and, hence, that
\begin{equation}
L'|_{\pi^{*}\cE}=L|_{\pi^{*}\cE} \otimes
\cO_{X}(\pi^{*}\cE)|_{\pi^{*}\cE}.
\label{eq:90}
\end{equation}
But
\begin{equation}
c_{1}(\cO_{X}(\pi^{*}\cE)|_{\pi^{*}\cE})=\pi^{*}\cE|_{\pi^{*}\cE}=
\pi^{*}\cE \cdot \pi^{*}\cE,
\label{eq:91}
\end{equation}
which vanishes using equations~\eqref{eq:16} and~\eqref{eq:23a}. It
follows that
\begin{equation}
\cO_{X}(\pi^{*}\cE)|_{\pi^{*}\cE}=\cO_{\pi^{*}\cE},
\label{eq:92}
\end{equation}
from which expression~\eqref{eq:88} follows. This is a result of
the fact that $\cE \cdot \cE=0$ and will not be true for the lift of curve
$\cS$.

It is useful to note that
\begin{equation}
\pi^{*}\cE=K3.
\label{eq:93}
\end{equation}
To prove this, recall that K3 is defined as a twofold elliptically fibered
over ${\mathbb P}^{1}$ with 24 singular fibers. 
Now consider $\pi^{*}\cE$, which
is indeed a twofold elliptically fibered over $\cE={\mathbb
P}^{1}$. How many singular fibers does it have? Note that the line bundle
of
the discriminant curve of base $B$ is
\begin{equation}
\Delta=-12K_{B},
\label{eq:94}
\end{equation}
where $K_{B}$ is the canonical bundle on $B$. It follows that
\begin{equation}
c_{1}(\Delta)=12c_{1}(B),
\label{eq:95}
\end{equation}
where we have used equation~\eqref{eq:21}. For $B={\mathbb F}_{r}$, we see
from~\eqref{eq:25} that the discriminant curve in ${\mathbb F}_{r}$ is
\begin{equation}
{\cal{D}}=24\cS+12(r+2)\cE.
\label{eq:96}
\end{equation}
Therefore, the number of discriminant points in $\cE={\mathbb P}^{1}$ is
\begin{equation}
{\cal{D}} \cdot \cE = 24 \cS \cdot \cE+12(r+2)\cE \cdot \cE=24,
\label{eq:97}
\end{equation}
where we have used~\eqref{eq:16}. It follows that $\pi^{*}\cE$ is a K3
surface. Henceforth, we will denote $\pi^{*}\cE$ by K3.

The short exact sequence~\eqref{eq:86} is now written as
\begin{equation}
0 \rightarrow L \rightarrow L' \rightarrow
L'|_{K3} \rightarrow 0.
\label{eq:98}
\end{equation}
This implies, in the usual way, a long exact sequence given in part by
\begin{equation}
0 \rightarrow H^{0}(X,L) \rightarrow H^{0}(X,L') \rightarrow
H^{0}(K3,L'|_{K3}) \rightarrow H^{1}(X,L) \rightarrow. 
\label{eq:99}
\end{equation}
Since $\cC$ is positive, it follows from the Kodaira vanishing
theorem~\eqref{eq:48} that
\begin{equation}
H^{1}(X,L)=0.
\label{eq:100}
\end{equation}
Hence
\begin{equation}
0 \rightarrow H^{0}(X,L) \rightarrow H^{0}(X,L') \rightarrow
H^{0}(K3,L'|_{K3}) \rightarrow 0,
\label{eq:101}
\end{equation}
which implies that
\begin{equation}
H^{0}(K3,L'|_{K3}) = H^{0}(X,L')/H^{0}(X,L).
\label{eq:102}
\end{equation}
Therefore
\begin{equation}
h^{0}(K3,L'|_{K3}) = h^{0}(X,L')-h^{0}(X,L)
\label{eq:103}
\end{equation}
and, hence
\begin{equation}
h^{0}(K3,L'|_{K3}) = n(V')-n(V)=n_{tm}.
\label{eq:104}
\end{equation}
We conclude that the transition moduli are precisely the moduli associated
with $\cO_{X}(\cC')|_{K3}$ or, equivalently, the restriction of spectral
cover
$\cC'$ to the lift of $z=\cE$, as claimed.

This result can be obtained by direct computation, as we now show. In this
context, the Riemann-Roch theorem states that
\begin{equation}
h^{0}(K3,L|_{K3})-h^{1}+h^{2}=\chi_{E}(K3,L|_{K3}),
\label{eq:105}
\end{equation}
where $\chi_{E}(K3,L|_{K3})$ is the Euler characteristic and $h^{q}$ for
$q=1,2$
are the dimensions of the higher cohomology groups of $K3$ evaluated in
$L|_{K3}$. To proceed, we must find the conditions under which the line
bundle
$L|_{K3}$ is positive. Recall from~\eqref{eq:87} and~\eqref{eq:87a} that
\begin{equation}
L=\cO_{X}(n\sigma+\pi^{*}(a\cS+b\cE))
\label{eq:106}
\end{equation}
and, hence, that
\begin{equation}
L|_{K3}=\cO_{K3}(n\sigma|_{K3}+a(\pi^{*}\cS)|_{K3}+b(\pi^{*}\cE)|_{K3}).
\label{eq:107}
\end{equation}
Using the facts that
\begin{equation}
\pi^{*}\cS|_{K3}=\pi^{*}\cS \cdot \pi^{*}\cE=\pi^{*}(\cS \cdot
\cE)=\pi^{*}(1)=F
\label{eq:108}
\end{equation}
and 
\begin{equation}
\pi^{*}\cE|_{K3}=\pi^{*}\cE \cdot \pi^{*}\cE=\pi^{*}(\cE \cdot \cE)=0,
\label{eq:109}
\end{equation}
where we have used~\eqref{eq:16} and~\eqref{eq:23a}, it follows that
\begin{equation}
L|_{K3}=\cO_{K3}(\cC|_{K3}),
\label{eq:110}
\end{equation}
where
\begin{equation}
\cC|_{K3}=n\sigma|_{K3}+aF.
\label{eq:111}
\end{equation}
We know from the discussion in Section 3 that $L|_{K3}$ is positive if and
only if $\cC|_{K3} \cdot c > 0$ for every effective curve $c$ in $K3$. It
is easy to see that the basis of effective curves in $K3$ is given by
\begin{equation}
\sigma|_{K3}, \qquad F.
\label{eq:112}
\end{equation}
First consider $\cC|_{K3} \cdot F$. It follows from the fact
\begin{equation}
\sigma \cdot F=1, \qquad F \cdot F=0
\label{eq:113}
\end{equation}
that
\begin{equation}
\cC|_{K3} \cdot F=n,
\label{eq:114}
\end{equation}
which is always positive. Now consider $\cC|_{K3} \cdot \sigma|_{K3}$. To
evaluate this, note, using~\eqref{eq:20e}, that
\begin{equation}
\sigma|_{K3} \cdot \sigma|_{K3}=-\pi^{*}(c_{1}({\mathbb
F}_{r})) \cdot \sigma \cdot \pi^{*}\cE.
\label{eq:115}
\end{equation}
Then, using~\eqref{eq:16},~\eqref{eq:23a},~\eqref{eq:25}
and~\eqref{eq:113} we
find that
\begin{equation}
\cC|_{K3} \cdot \sigma|_{K3}= -2n+a,
\label{eq:116}
\end{equation}
which is positive if and only if 
\begin{equation}
a > 2n.
\label{eq:117}
\end{equation}
However, the positivity of $\cC$ already demands that $a>2n$, as stated
in~\eqref{eq:72}. Therefore, we conclude that $L|_{K3}$ and, hence, 
$\cC|_{K3}$ is positive. This allows us to 
employ the Kodaira vanishing theorem
to evaluate the higher cohomology groups. A straightforward extension of
the
discussion in Section 4 allows us to conclude that
\begin{equation}
H^{q}(K3,L|_{K3})=0
\label{eq:118}
\end{equation}
for $q>0$. It follows that $h^{q}=0$ for $q=1,2$ and the Riemann-Roch
theorem~\eqref{eq:105} simplifies to
\begin{equation}
h^{0}(K3, L|_{K3})=\chi_{E}(K3, L|_{K3}).
\label{eq:119}
\end{equation}
The Euler characteristic is determined from the Atiyah-Singer index
theorem to be
\begin{equation}
\chi_{E}(K3, L|_{K3})=\int_{K3}{ ch(\cO_{K3}(\cC|_{K3})) \wedge Td(TK3) },
\label{eq:120}
\end{equation}
which, on the Calabi-Yau twofold $K3$ takes the form
\begin{equation}
\chi_{E}(K3, L|_{K3})=\frac{1}{2}\int_{K3}{
c_{1}^{2}(\cO_{K3}(\cC|_{K3}))}
+\frac{1}{12}\int_{K3}{ c_{2}(TK3)}.
\label{eq:121}
\end{equation}
Then, using~\eqref{eq:111} and the fact that
\begin{equation}
c_{2}(TK3)=24,
\label{eq:122}
\end{equation}
we find 
\begin{equation}
h^{0}(K3, L|_{K3})=2+an-n^{2}.
\label{eq:123}
\end{equation}
We conclude by noting that for $a'=a$ and $b'=b+1$, equation~\eqref{eq:80}
becomes
\begin{equation}
n_{tm}=2+an-n^{2}.
\label{eq:124}
\end{equation}
It follows from this and the relation~\eqref{eq:88} that
\begin{equation}
h^{0}(K3, L'|_{K3})=n_{tm},
\label{eq:125}
\end{equation}
as in~\eqref{eq:104}. 

Having proven the claim for $z=\cE$, we now extend our result
to small instanton transitions with curve $z=\cS$. The 
exact sequence of bundles~\eqref{eq:98} now reads
\begin{equation}
0 \rightarrow L \rightarrow L' \rightarrow
L'|_{\pi^{*}\cS} \rightarrow 0,
\label{eq:E1}
\end{equation}
where
\begin{equation}
L=\cO_{X}(\cC), \qquad L^{\prime}=\cO_{X}(\cC')
\label{eq:E2}
\end{equation}
and
\begin{equation}
\cC=n\sigma + \pi^{*}(a\cS+b\cE), \qquad \cC'=n\sigma
+\pi^{*}((a+1)\cS+b\cE).
\label{eq:E2a}
\end{equation}

Information about the surface $\pi^{*}\cS$, elliptically fibered over $\cS
={\mathbb P}^{1}$,
can be obtained by counting the number of degenerate fibers. As explained
above, this number can be found by intersecting the discriminant curve 
${\cal{D}}$ given in equation~\eqref{eq:96}
with the curve $\cS$ on the base. Using equation~\eqref{eq:16},
one finds
\begin{equation}
{\cal{D}}\cdot \cS = 12 (2-r).
\label{eq:E3}
\end{equation}
When $r=0$, we see that the surface $\pi^{*}\cS$
has 24 degenerate fibers implying that $\pi^{*}\cS=K3$. For $r=1$,
the number of degenerate fibers is 12 and, hence,  $\pi^{*}\cS$ is the
del Pezzo surface $dP_9$, while for $r=2$ the resulting surface is a
nonsingular
fibration. One can show that this fibration is simply the cross product
${\mathbb P}^{1} \times T^2$.

As before, the exact sequence of bundles~\eqref{eq:E1} produces the exact 
sequence of cohomology groups 
\begin{equation}
0 \rightarrow H^{0}(X,L) \rightarrow H^{0}(X,L') \rightarrow
H^{0}(\pi^{*}\cS,L'|_{\pi^{*}\cS}) \rightarrow 0,
\label{eq:E4}
\end{equation}
where we have used the vanishing of $H^1(X,L)$ discussed above. Therefore 
\begin{equation}
h^0(\pi^{*}\cS, L^{\prime}|_{\pi^{*}\cS})= h^0(X,L^{\prime})-
h^{0}(X,L)
\label{eq:E5}
\end{equation}
and, hence
\begin{equation}
h^0(\pi^{*}\cS, L^{\prime}|_{\pi^{*}\cS})= n(V^{\prime})-n(V)=n_{tm}.
\label{eq:E6}
\end{equation}
In analogy with the previous case, we see that the transition moduli are
associated 
with the moduli of the bundle $\cO_{X}(\cC')|_{\pi^{*}\cS}$ or,
equivalently,
the restriction of the spectral cover $\cC'$ to the lift of the 
$z=\cS$ curve.

For completeness, we now compute
$h^0(\pi^{*}\cS, L^{\prime}|_{\pi^{*}\cS})$
by an independent calculation and demonstrate that the result is identical
to $n_{tm}$ found from the general formula~\eqref{eq:80}. 
Using equation~\eqref{eq:16}, we get
\begin{equation}
\pi^{*}\cS |_{\pi^{*}\cS} =\pi^{*}(\cS \cdot \cS) = -rF, \qquad
\pi^{*}\cE  |_{\pi^{*}\cS} =\pi^{*}(\cE \cdot \cS) = F
\label{eq:E7}
\end{equation}
and, therefore, the bundle $L^{\prime}$ restricted to $\pi^{*}\cS$
is given by
\begin{equation}
L^{\prime}|_{\pi^{*}\cS} = \cO_{\pi^{*}\cS}(\cC'|_{\pi^{*}\cS}),
\label{eq:E8}
\end{equation}
where 
\begin{equation}
\cC'|_{\pi^{*}\cS}=n\sigma|_{\pi^{*}\cS}+(b-(a+1)r)F.
\label{eq:E8a}
\end{equation}
Here $ \sigma |_{\pi^{*}\cS}$ represents the global section of the 
$\pi^{*}\cS$ elliptic fibration. As before, we would like to use the
Riemann-Roch theorem to compute $h^0(\pi^{*}\cS,
L^{\prime}|_{\pi^{*}\cS})$.
From the previous analysis, it follows that for this to be straightforward
the
higher cohomology groups have to vanish
\begin{equation}
H^{q}(\pi^{*}\cS, L^{\prime}|_{\pi^{*}\cS})=0, \qquad q>0.
\label{eq:E9}
\end{equation}
This last equation is guaranteed by the Kodaira vanishing 
theorem~\eqref{eq:44} provided the bundle $L^{\prime}$ restricted to 
$\pi^{*}\cS$ can be represented as
\begin{equation}
L^{\prime} |_{\pi^{*}\cS}= K_{\pi^{*}\cS}\otimes 
L^{\prime \prime}_{\pi^{*}\cS},
\label{eq:E10}
\end{equation}
where $K_{\pi^{*}\cS}$ is the canonical bundle of $\pi^{*}\cS $
and $L^{\prime \prime}_{\pi^{*}\cS}$ is some positive line bundle.
Note that, unlike the $z=\cE$ case, surface $\pi^{*}\cS $
is not a Calabi-Yau twofold (unless r=0) and, hence, the canonical bundle 
$ K_{\pi^{*}\cS}$ is nontrivial. In fact, using the adjunction formula we
see
that
\begin{equation}
K_{\pi^{*}\cS}=K_X \otimes \cO_{X}(\pi^{*}\cS)|_{\pi^{*}\cS}=
\cO_{X}(\pi^{*}\cS) |_{\pi^{*}\cS},
\label{eq:E11}
\end{equation}
where we have used that fact that $K_{X}$ is trivial. But
\begin{equation}
c_{1}(\cO_{X}(\pi^{*}\cS) |_{\pi^{*}\cS})=\pi^{*}(\cS \cdot \cS)=-rF.
\label{eq:E11a}
\end{equation}
It follows that
\begin {equation}
K_{\pi^{*}\cS}=\cO_{\pi^{*}\cS}(-rF).
\label{eq:E11b}
\end{equation}
Using equations~\eqref{eq:E8},~\eqref{eq:E8a},~\eqref{eq:E10}
and~\eqref{eq:E11b}, 
we conclude that the line 
bundle $L^{\prime \prime}_{\pi^{*}\cS}$ is of the form
\begin{equation}
L^{\prime \prime}_{\pi^{*}\cS} = 
\cO_{\pi^{*}\cS}(\cC''_{\pi^{*}\cS}),
\label{eq:E12}
\end{equation}
where
\begin{equation}
\cC''_{\pi^{*}\cS}=n\sigma|_{\pi^{*}\cS}+(b-ar)F.
\label{eq:E12a}
\end{equation}
Let us find under what conditions $L^{\prime \prime}_{\pi^{*}\cS}$ is
positive. As before, we have to require that
$\cC''_{\pi^{*}\cS} \cdot c > 0$
for all effective curves $c$ which, in this case, have the basis
\begin{equation}
\sigma|_{\pi^{*}\cS}, \qquad F.
\label{eq:E13a}
\end{equation}
By using relations~\eqref{eq:20e},~\eqref{eq:113} and the fact that
\begin{equation}
\sigma |_{\pi^{*}\cS} \cdot \sigma |_{\pi^{*}\cS} =-
\pi^{*}(c_1 ( {\mathbb F}_{r})) \cdot \sigma \cdot
\pi^{*}\cS = -(2-r),
\label{eq:E14}
\end{equation}
we find that the line bundle 
$L^{\prime \prime}_{\pi^{*}\cS}$ is positive if and only if
\begin{equation}
b> ar + (2-r) n.
\label{eq:E15}
\end{equation}
But condition~\eqref{eq:E15} is just the positivity
condition~\eqref{eq:26a}.
Therefore, if the original spectral cover $\cC$ is chosen to be positive
the line bundle $L^{\prime \prime}_{\pi^{*}\cS}$ is automatically
positive. As a consequence, equation~\eqref{eq:E9} is valid and 
$h^{0} (\pi^{*}\cS, L^{\prime} |_{\pi^{*}\cS} )$ coincides with
the Euler characteristic
\begin{equation}
h^{0} (\pi^{*}\cS, L^{\prime} |_{\pi^{*}\cS} )=
\chi_{E} (\pi^{*}\cS, L^{\prime} |_{\pi^{*}\cS} ).
\label{eq:E17}
\end{equation}
The latter can be computed by using the index theorem
\begin{equation}
\chi_{E} (\pi^{*}\cS, L^{\prime} |_{\pi^{*}\cS} ) =
\int_{\pi^{*}\cS} ch(\cO_{\pi^{*}\cS}
( \cC'|_{\pi^{*}\cS})) \wedge Td(T\pi^{*}\cS)
\label{eq:E18}
\end{equation}
which, on the twofold $\pi^{*}\cS$, takes the form
\begin{eqnarray} 
\chi_{E} (\pi^{*}\cS, L'_{\pi^{*}\cS}) & = &
\frac{1}{2} \int_{\pi^{*}\cS} c_1^2 
(\cO_{\pi^{*}\cS}( \cC^{\prime} |_{\pi^{*}\cS})) +
\frac{1}{2} \int_{\pi^{*}\cS} 
c_1(\cO_{\pi^{*}\cS}( \cC^{\prime} |_{\pi^{*}\cS})) \wedge
c_1 (T\pi^{*}\cS) \nonumber \\
 & & +\frac{1}{12}\int_{\pi^{*}\cS}(c_2(T\pi^{*}\cS) +c_1^2(T\pi^{*}\cS)). 
\label{eq:E18a}
\end{eqnarray}
The first Chern class, $c_1(T\pi^{*}\cS)$, is given by 
\begin{equation}
c_1(T\pi^{*}\cS)= -c_1(K_{\pi^{*}\cS})=rF,
\label{eq:E19}
\end{equation}
where we have used~\eqref{eq:E11}. Note that
\begin{equation}
c_1^2(T\pi^{*}\cS)=0.
\label{eq:E20}
\end{equation}
Finding $c_2(T\pi^{*}\cS)$ requires some work. Consider the short exact 
sequence 
\begin{equation}
0 \rightarrow T\pi^{*}\cS \rightarrow  TX|_{\pi^{*}\cS}  
\rightarrow N\pi^{*}\cS \rightarrow 0,
\label{eq:E21}
\end{equation}
where $N\pi^{*}\cS$ is the normal bundle. The sequence~\eqref{eq:E21}
follows directly from the definition of the normal bundle
\begin{equation}
N\pi^{*}\cS = TX|_{\pi^{*}\cS}{/} T\pi^{*}\cS.
\label{eq:E22}
\end{equation}
On the other hand, we can find the normal bundle by relating it to the
line
bundle $\cO_{X}(\pi^{*}\cS)|_{\pi^{*}\cS}$ using
\begin{equation}
N\pi^{*}\cS = \cO_{X}(\pi^{*}\cS)|_{\pi^{*}\cS} = K_{\pi^{*}\cS}=
\cO_{\pi^{*}\cS}(-rF),
\label{eq:E23}
\end{equation}
where equations~\eqref{eq:E11} and~\eqref{eq:E11b}
have been used. Exact sequence~\eqref{eq:E21}
then produces the following relation for total Chern classes
\begin{equation}
c( TX|_{\pi^{*}\cS})= c( T\pi^{*}\cS) \wedge 
c(\cO_{\pi^{*}\cS}(-rF)).
\label{eq:E24}
\end{equation}
In particular, we have
\begin{equation}
c_2 ( T\pi^{*}\cS) = c_2 (TX|_{\pi^{*}\cS}).
\label{eq:E25}
\end{equation}
Using expression~\eqref{eq:50b} for $c_2(TX)$, we conclude that
\begin{equation}
c_2(T\pi^{*}\cS) = 12 \sigma \cdot \pi^{*}(c_1 ({\mathbb F}_r )) \cdot
\pi^{*}\cS = 12 (2-r).
\label{eq:E26}
\end{equation}
One can now calculate the Euler characteristic~\eqref{eq:E18a}
and, therefore, $h^{0} (\pi^{*}\cS, L^{\prime} |_{\pi^{*}\cS} )$.
The result is
\begin{equation}
h^{0} (\pi^{*}\cS, L^{\prime} |_{\pi^{*}\cS} ) = 
nb- (n^2 - 2) +
\frac{rn}{2} (n-1) -r (na+1).
\label{eq:E27}
\end{equation}
On the other hand, equation~\eqref{eq:80} with $a^{\prime} =a+1$,
$b^{\prime}=b$ becomes
\begin{equation}
n_{tm}=
nb- (n^2 - 2) +
\frac{rn}{2} (n-1) -r (na+1).
\label{eq:E28}
\end{equation}
We see that equations~\eqref{eq:E27} and~\eqref{eq:E28} coincide and,
hence
that
\begin{equation}
h^{0}(\pi^{*}\cS, L'|_{\pi^{*}\cS})=n_{tm}
\label{eq:E28a}
\end{equation}
as in~\eqref{eq:E6}.\\

\noindent Using~\eqref{eq:104} and~\eqref{eq:E6}, 
it is now straightforward to prove the following.

\begin{itemize}

\item The transition moduli arise as the moduli of the spectral cover
$\cC'$
restricted to the lift, $\pi^{*}z$, of the horizontal curve $z$, for any
effective curve $z$. Specifically,
\begin{equation}
h^{0}(\pi^{*}z, L'|_{\pi^{*}z})=n_{tm},
\label{eq:E28b}
\end{equation}
where $n_{tm}$ is given in expression~\eqref{eq:80}.

\end{itemize}

\noindent To be concrete, let us give several examples.\\

\noindent $\bf Example \rm$ $\bf 1: \rm$ Consider the small instanton
transition presented in Example 1 earlier in this Section. Recall that the
pre-transition vector bundle $V$ was specified by $B={\mathbb F}_{1}$,
$G=SU(3)$, $a=7$, $b=12$ and $\lambda=1/2$. In this case, the
$M$5-brane involved in the transition is wrapped on the curve
\begin{equation}
z=\cE,
\label{eq:E28c}
\end{equation}
$\pi^{*}\cE=K3$ and
\begin{equation}
L'|_{K3}=\cO_{K3}(3\sigma|_{K3}+7F),
\label{eq:E28d}
\end{equation}
where we have used~\eqref{eq:111}. Then, we conclude from~\eqref{eq:80c} 
and~\eqref{eq:E28b} that
\begin{equation}
h^{0}(K3, L'|_{K3})=14
\label{eq:E28e}
\end{equation}
is the number of transition moduli.\\

\noindent $\bf Example \rm$ $\bf 2: \rm$ Consider the small instanton
transition presented in Example 2 earlier in this Section. Recall that the
pre-transition vector bundle $V$ was again specified by $B={\mathbb
F}_{1}$,
$G=SU(3)$, $a=7$, $b=12$ and $\lambda=1/2$. In this case, however, the
$M$5-brane involved in the transition is wrapped on the curve
\begin{equation}
z=\cS,
\label{eq:E28cc}
\end{equation}
$\pi^{*}\cS=dP_{9}$ and
\begin{equation}
L'|_{dP_{9}}=\cO_{dP_{9}}(3\sigma|_{dP_{9}}+4F),
\label{eq:E28dd}
\end{equation}
where we have used~\eqref{eq:E8a}. We conclude from~\eqref{eq:80cc} 
and~\eqref{eq:E28b} that
\begin{equation}
h^{0}(dP_{9}, L'|_{dP_{9}})=10
\label{eq:E28ee}
\end{equation}
is the number of transition moduli.


\section{Evaluation of Transition Moduli on the Curve $z$:}


In the previous section, we have presented an interpretation of the
transition moduli as the moduli of the spectral cover $\cC'$ restricted to
$\pi^{*}z$. However, it is clearly of interest to ask whether the
transition moduli can be related directly to intrinsic quantities on the 
curve $\pi^{*}z \cdot \sigma$ alone. The answer to this is in
the affirmative, as we will now demonstrate.

As above, we begin by restricting the discussion to the horizontal curve
$z=\cE$. Recall from~\eqref{eq:88} and~\eqref{eq:110}, 
~\eqref{eq:111} that the line bundle associated with the
transition moduli is $L'|_{K3}=L|_{K3}$, which we now denote by 
$\cL_{a}$ to streamline our notation. That is
\begin{equation}
\cL_{a}=\cO_{K3}(n\sigma|_{K3}+aF).
\label{eq:126}
\end{equation}
Note that this can be written as the tensor product
\begin{equation}
\cL_{a}=\cL_{0} \otimes \pi^{*}\cO_{\cE}(a),
\label{eq:126a}
\end{equation}
where 
\begin{equation} 
\cL_{0}=\cO_{K3}(n\sigma|_{K3}).
\label{eq:127}
\end{equation}
We now want to consider the push-forward of the line bundle $\cL_{a}$ 
onto the curve $z=\cE$ given by
\begin{equation}
\pi_{*}\cL_{a}=\pi_{*}(\cL_{0} \otimes \pi^{*}\cO_{\cE}(a)).
\label{eq:128}
\end{equation}
Using the projection formula 
\begin{equation}
\pi_{*}(\cF \otimes \pi^{*}{\cal{G}})=(\pi_{*}\cF) \otimes {\cal{G}}
\label{eq:129}
\end{equation}
for arbitrary bundles $\cF$ and ${\cal{G}}$, equation~\eqref{eq:128}
becomes
\begin{equation}
\pi_{*}\cL_{a}=\pi_{*}\cL_{0} \otimes \cO_{\cE}(a).
\label{eq:130}
\end{equation}
Now, since the spectral cover $\cC'|_{K3}$ 
is an $n$-fold cover of the base curve $\cE$, it follows that the
direct image of
any line bundle on $\cC'|_{K3}$ onto $\cE$ is a rank $n$ vector bundle. At
this point, for specificity, it is convenient to choose a value of $n$,
which
we take to be
\begin{equation}
n=3.
\label{eq:140}
\end{equation}
Our remarks will remain true for any value of $n$. It follows that we can
always represent $\pi_{*}\cL_{0}$ as
\begin{equation}
\pi_{*}\cL_{0}=\cO_{\cE}(A) \oplus \cO_{\cE}(B) \oplus \cO_{\cE}(C)
\label{eq:141}
\end{equation}
for some integers $A,B$ and $C$. Then from~\eqref{eq:130} we have 
\begin{equation}
\pi_{*}\cL_{a}=(\cO_{\cE}(A) \oplus \cO_{\cE}(B) \oplus \cO_{\cE}(C))
\otimes
\cO_{\cE}(a)
\label{eq:142}
\end{equation}
and, hence, that
\begin{equation}
\pi_{*}\cL_{a}=\cO_{\cE}(a+A) \oplus \cO_{\cE}(a+B) \oplus \cO_{\cE}(a+C).
\label{eq:143}
\end{equation}
We now need to compute the values of integers $A,B$ and $C$.

One can show, using a Leray spectral sequence, 
that for any value of $n$ and $a$
\begin{equation}
H^{0}(\cE, \pi_{*}\cL_{a})=H^{0}(K3, \cL_{a}).
\label{eq:144}
\end{equation}
Here, we will continue to assume that $n=3$. Let us first consider the
case 
where $a=0$ and, therefore, $\cL_{a}=\cL_{0}=\cO_{K3}(3\sigma|_{K3})$.
Now, it
follows
from~\eqref{eq:116} that $\sigma|_{K3}^{2}=-2$ and, hence, that the curve
$3\sigma|_{K3}$ is isolated. Therefore,
\begin{equation}
h^{0}(K3, \cL_{0})=1,
\label{eq:145}
\end{equation}
which implies, using~\eqref{eq:144}, that
\begin{equation}
h^{0}(\cE, \pi_{*}\cL_{0})=1.
\label{eq:146}
\end{equation}
However, line bundles $\cO_{{\mathbb P}^{1}}(m)$ over a projective line
${\mathbb P}^{1}$ have the generic property that
\begin{equation}
dimH^{0}({\mathbb P}^{1}, \cO_{{\mathbb P}^{1}}(m))= m+1
\label{eq:147}
\end{equation}
for $m \geq 0$ and
\begin{equation}
dimH^{0}({\mathbb P}^{1}, \cO_{{\mathbb P}^{1}}(m))=0
\label{eq:148}
\end{equation}
for $m < 0$. Therefore, putting
together~\eqref{eq:141},~\eqref{eq:146},~\eqref{eq:147} and~\eqref{eq:148} 
we see, without loss of generality, that we can choose
\begin{equation}
A=0, \qquad B,C < 0.
\label{eq:149}
\end{equation}
Second, let us consider the case where $a > 6$. Since, for $n=3$, this
satisfies the positivity condition~\eqref{eq:117}, then it follows
from~\eqref{eq:123} and ~\eqref{eq:144} that
\begin{equation}
h^{0}( \cE, \pi_{*}\cL_{a})=3a-7.
\label{eq:150}
\end{equation}
The reader can easily verify that, since $A,B$ and $C$ are fixed
coefficients,
one can only solve~\eqref{eq:150} for arbitrary values of $a > 6$ if
\begin{equation}
a+B \geq 0, \qquad a+C \geq 0.
\label{eq:151}
\end{equation}
It then follows from~\eqref{eq:143} and~\eqref{eq:147} that
\begin{equation}
h^{0}(\cE, \pi_{*}\cL_{a})= (a+1)+(a+B+1)+(a+C+1).
\label{eq:152}
\end{equation}
But, from~\eqref{eq:150} this must equal $3a-7$, from which we find that
\begin{equation} 
B+C=-10.
\label{eq:153}
\end{equation}
Third, let us now find the formula for $h^{0}(K3, \cL_{a})$, and, hence,
using
~\eqref{eq:144} for $h^{0}(\cE, \pi_{*}\cL_{a})$, 
for the remaining values of $a$, $0<a \leq 6$. Note that $\sigma|_{K3}$ is
a
divisor in $K3$. Let ${\cal{D}} \subset K3$ be any divisor and 
$E=\cO_{K3}({\cal{D}})$ the
associated line bundle. Then, it follows from~\eqref{eq:81} that there is
the
short exact sequence
\begin{equation}
0 \rightarrow \cO_{K3} ({\cal{D}}-\sigma|_{K3}) \rightarrow
\cO_{K3}({\cal{D}}) \rightarrow \cO_{\sigma|_{K3}}({\cal{D}} 
\cdot \sigma|_{K3}) \rightarrow 0,
\label{eq:I0}
\end{equation}
where we have used ${\cal{D}}|_{K3}={\cal{D}} \cdot \sigma|_{K3}$. 
This then implies the long exact sequence of cohomology groups given by
\begin{eqnarray}
&& 0 \rightarrow H^0 (K3, \cO_{K3} ({\cal{D}}-\sigma|_{K3})) 
\rightarrow H^0 (K3, \cO_{K3} ({\cal{D}}))
\rightarrow  H^0 ({\mathbb P}^1, \cO_{{\mathbb P}^1} 
({\cal{D}} \cdot\sigma|_{K3})) \nonumber \\
&&\rightarrow H^1 (K3, \cO_{K3} ({\cal{D}} -\sigma|_{K3})) \rightarrow.
\label{eq:I1}
\end{eqnarray}
In the appropriate places, we have inserted the fact that
$\sigma|_{K3}={\mathbb P}^{1}$. First, assume that ${\cal{D}} \cdot
\sigma|_{K3} <
0$. Then, it follows from~\eqref{eq:148} 
that $h^{0}({\mathbb P}^{1}, \cO_{{\mathbb
P}^{1}}({\cal{D}} \cdot \sigma|_{K3}))=0$ and, hence
\begin{equation}
h^0 (K3, \cO_{K3} ({\cal{D}})) = h^0 (K3,
\cO_{K3}({\cal{D}}-\sigma|_{K3})),
\qquad  {\cal{D}}\cdot \sigma|_{K3} < 0. 
\label{eq:I2} 
\end{equation}
Now assume ${\cal{D}} \cdot \sigma|_{K3} \geq 0$. Equation~\eqref{eq:147} 
then implies 
\begin{equation}
h^{0}({\mathbb P}^{1}, \cO_{{\mathbb P}^{1}}({\cal{D}} \cdot
\sigma|_{K3}))=
{\cal{D}} \cdot \sigma|_{K3}+1.
\label{eq:I2a}
\end{equation}
Note, using $\sigma|_{K3}^{2}=-2$, that $({\cal{D}}-\sigma|_{K3}) \cdot
\sigma|_{K3}
>0$. If we further assume that ${\cal{D}} \cdot F >1$, then $({\cal{D}}
 -\sigma|_{K3}) \cdot F >0$ and, hence, ${\cal{D}} -\sigma|_{K3}$ 
is a positive divisor in $K3$. It
then follows from the Kodaira vanishing theorem that
\begin{equation}
H^{q}(K3,\cO_{K3}({\cal{D}} -\sigma|_{K3}))=0
\label{eq:I2b}
\end{equation}
for $q > 0$. Hence, the exact sequence~\eqref{eq:I1} truncates to 
\begin{equation}
0 \rightarrow H^0 (K3, \cO_{K3} ({\cal{D}}-\sigma|_{K3})) 
\rightarrow H^0 (K3, \cO_{K3} ({\cal{D}}))
\rightarrow  H^0 ({\mathbb P}^1, \cO_{{\mathbb P}^1} 
({\cal{D}} \cdot\sigma|_{K3})) \rightarrow 0. 
\label{eq:I2c}
\end{equation}
Therefore, using~\eqref{eq:I2a}, for ${\cal{D}}-\sigma|_{K3} > 0$ we have 
\begin{equation}
h^0 (K3, \cO_{K3} ({\cal{D}})) - h^0 (K3, \cO_{K3}
({\cal{D}}-\sigma|_{K3})) =
{\cal{D}}\cdot \sigma|_{K3} +1, \qquad  {\cal{D}}  \cdot \sigma|_{K3} \geq
0. 
\label{eq:I3}
\end{equation}
Now, from equations~\eqref{eq:I2} and~\eqref{eq:I3} we can compute 
$h^0 (K3, \cO_{K3} (k \sigma|_{K3} +a F))$ for $k=0,1,2,3$.
Start with $k=0$. In this case, equations~\eqref{eq:144}
and~\eqref{eq:147} imply
that
\begin{equation}
h^0(K3, \cO_{K3} (aF)) = h^0(K3, \pi^{*}\cO_{\cE}(a))=
h^0(\cE, \cO_{\cE}(a)) = a+1, \qquad a \geq 0.
\label{eq:I5}
\end{equation}
Now take $k=1$. Any effective divisor of the class 
$\sigma|_{K3} + aF$ has intersection 
number 1 with $F$ and, hence, consists of some section $\sigma'$ of $K3$
plus
fibers. A special property of elliptically fibered $K3$ twofolds is that
every section $\sigma'$ satisfies $(\sigma')^{2}=-2$ and, therefore, is
rigid.
Thus, every point in the projective space ${\mathbb
P}H^{0}(K3, \cO_{K3}(\sigma|_{K3}+aF))$ corresponds  to some $\sigma'$
plus
fibers, and none of the $\sigma'$ can move continuously. Therefore, all
the 
$\sigma'$ must be equal to one another and, hence, to the rigid section
$\sigma|_{K3}$. We conclude that the only freedom to move the divisor is
in
the fibers, so we get 
\begin{equation}
h^0 (K3, \cO_{K3} (\sigma|_{K3} +a F))=
h^0 (K3, \cO_{K3} (aF)) =a+1, \qquad a \geq 0.
\label{eq:I6}
\end{equation}  
As the next step, take $k=2$ and set ${\cal{D}}=2\sigma|_{K3}+aF$. 
First, note that
\begin{equation}
(2\sigma|_{K3}+aF) \cdot \sigma|_{K3} =a-4.
\label{eq:I6a}
\end{equation}
This breaks into two cases. For $0 \leq a \leq 3$, ${\cal{D}} \cdot 
\sigma|_{K3}<0$. Then, it follows from~\eqref{eq:I2} and~\eqref{eq:I6}
that
\begin{equation}
h^0 (K3, \cO_{K3} (2\sigma|_{K3} +a F))=
h^0 (K3, \cO_{K3} (\sigma|_{K3} +aF)) =a+1, \qquad 0 \leq  a \leq 3 .
\label{eq:I7}
\end{equation}  
Now let $a \geq 4$. In this case, ${\cal{D}} \cdot \sigma|_{K3} \geq 0$
and
\begin{equation}
(\sigma|_{K3}+aF) \cdot \sigma|_{K3} =a-2 >0, \qquad 
(\sigma|_{K3}+aF) \cdot F=1 >0.
\label{eq:I7a}
\end{equation}
Then equation~\eqref{eq:I3} is applicable and implies that
\begin{equation}
h^0 (K3, \cO_{K3} (2\sigma|_{K3} +a F))=
h^0 (K3, \cO_{K3} (\sigma|_{K3} +aF)) +a-3 =2a-2 , \qquad   a \geq 4 
\label{eq:I8}
\end{equation}  
where we have used~\eqref{eq:I6}. Finally, consider $k=3$
and set ${\cal{D}}=3\sigma|_{K3}+aF$. Then
\begin{equation}
(3\sigma|_{K3}+aF) \cdot \sigma|_{K3}=a-6.
\label{eq:I8a}
\end{equation}
Hence, for $0 \leq a \leq 5$, ${\cal{D}} \cdot \sigma|_{K3} < 0$ and
equation~\eqref{eq:I2} implies that
\begin{equation}
h^0 (K3, \cO_{K3} (3\sigma|_{K3} +a F))=
h^0 (K3, \cO_{K3} (2\sigma|_{K3} +aF)) , \qquad 0 \leq  a  \leq 5. 
\label{eq:I9}
\end{equation}  
For $a \geq 6$, ${\cal{D}} \cdot \sigma|_{K3} \geq 0$ and 
\begin{equation}
(2\sigma|_{K3} +aF) \cdot \sigma|_{K3}=a-4 > 0, \qquad 
(2\sigma|_{K3} +aF) \cdot F=2 > 0.
\label{eq:I9a}
\end{equation}
Then, equation~\eqref{eq:I3} is applicable and gives
\begin{equation}
h^0 (K3, \cO_{K3} (3\sigma|_{K3} +a F))=
h^0 (K3, \cO_{K3} (2\sigma|_{K3} +aF)) + a-5 , \qquad  a  \geq 6. 
\label{eq:I10}
\end{equation}  
To get the required information about the coefficients
$B$ and $C$, it is sufficient to look at one particular value of $a$,
namely $a=4$. In this case, expressions~\eqref{eq:I8} and~\eqref{eq:I9}
tell us
that
\begin{equation}
h^0 (K3, \cO_{K3} (3\sigma|_{K3} +a F)) = 6.
\label{eq:I11}
\end{equation}
By using~\eqref{eq:143}, ~\eqref{eq:144},~\eqref{eq:149},
and~\eqref{eq:I11}, we conclude 
\begin{equation}
h^{0}(\cE, \pi_{*}\cL_{a} )=h^0 (\cE, \cO_{\cE}(4) \oplus
\cO_{\cE}(4+B) \oplus \cO_{\cE}(4+C))=6.
\label{eq:I12}
\end{equation}
Since the line bundle $\cO_{\cE}(4)$ already has $5$ sections, one of the
other two line bundles has to be trivial and one has to be negative.
Without loss of generality, one can choose
\begin{equation}
B=-4, \qquad C \leq -5
\label{eq:I12a} 
\end{equation}
as the solution of equation~\eqref{eq:I12}. Combining this result
with~\eqref{eq:153}, we see that
\begin{equation}
B=-4, \qquad C=-6.
\label{eq:154}
\end{equation}
Putting everything together, we conclude that for $n=3$ and arbitrary
coefficient $a$, the push-forward of line bundle $\cL_{a}$ onto curve
$\cE$ 
and, therefore, onto the curve $\pi^{*}\cE \cdot \sigma \subset K3$ 
is given by the rank three vector bundle 
\begin{equation}
\pi_{*}\cL_{a}=\cO_{\cE}(a) \oplus \cO_{\cE}(a-4) \oplus \cO_{\cE}(a-6).   
\label{eq:155}
\end{equation}
The sections of this vector bundle, specified by
\begin{equation}
h^{0}(\cE, \pi_{*}\cL_{a})=3a-7
\label{eq:156}
\end{equation}
are precisely the transition moduli of the associated small instanton
transition. It is useful at this point to give a concrete example.\\

\noindent $\bf Example: \rm$ Consider the small instanton transition
presented
in Example 1 earlier in this Section. Recall that the pre-transition
vector
bundle $V$ was specified by $B={\mathbb F}_{1}$, $G=SU(3)$, $a=7$, $b=12$
and
$\lambda=1/2$. In this case, the $M$5-brane involved in the transition is
wrapped on the curve
\begin{equation}
z=\cE.
\label{eq:156a}
\end{equation}
Then, from~\eqref{eq:155} we find that the transition moduli are the
holomorphic
sections of the rank $3$ vector bundle
\begin{equation}
\pi_{*}\cL_{7}=\cO_{\cE}(7) \oplus \cO_{\cE}(3) \oplus \cO_{\cE}(1) 
\label{eq:156b}
\end{equation}
over $z=\cE$, where, from~\eqref{eq:126}
\begin{equation}
\cL_{7}=\cO_{K3}(3\sigma|_{K3}+7F).
\label{eq:156c}
\end{equation}
It follows from~\eqref{eq:156} that the number of transition moduli is
\begin{equation}
h^{0}(\cE,\pi_{*}\cL_{7})=14,
\label{eq:156d}
\end{equation}
in agreement with~\eqref{eq:80c} and~\eqref{eq:E28e}.\\

Now consider the case $z=\cS$. In the previous section, we showed that
the transition moduli are given by the number of global sections of the
line 
bundle $L'|_{\pi^{*}\cS}$ specified in~\eqref{eq:E8} and~\eqref{eq:E8a}.
As before, denote
\begin{equation}
\cL_0 = \cO_{\pi^{*}\cS}(n\sigma |_{\pi^{*}\cS})
\label{eq:E01}
\end{equation}
and
\begin{equation}
\cL_{\gamma} =\cO_{\pi^{*}\cS}(n\sigma |_{\pi^{*}\cS} + \gamma F),
\qquad \gamma= b-(a+1)r.
\label{eq:E02}
\end{equation}
The line bundle $\cL_{\gamma}$ can be written as
\begin{equation}
\cL_{\gamma} =\cO_{\pi^{*}\cS}(n\sigma |_{\pi^{*}\cS}) 
\otimes  \pi^{*}\cO_{\cE}(\gamma)
\label{eq:E02a}
\end{equation}
and, by using projection formula~\eqref{eq:129}, we find that its direct 
image is given by
\begin{equation}
\pi_{*}\cL_{\gamma} = \pi_{*}\cL_0 \otimes \cO_{\cS} (\gamma).
\label{eq:E03}
\end{equation}
Note that since $\cC'|_{\pi^{*}\cS}$ is an $n$-fold cover of the base
curve
$\cS$, the push-forward of any line bundle of $\cC'|_{\pi^{*}\cS}$ onto
$\cS$
is
a rank $n$ vector bundle.

At this point, for concreteness, we specify
\begin{equation}
n=3, \qquad r=1,
\label{eq:E07}
\end{equation}
but leave coefficients $a$ and $b$ arbitrary. Our remarks, however, will
remain true for any values of $n$ and $r$. We remind the reader that 
when $r=1$, the surface $\pi^{*}\cS$ is the del Pezzo surface $dP_9$.

Since every line bundle on $\cS = {\mathbb P}^1$ is a sum of line bundles,
rank $3$ vector bundle $\pi_{*} \cL_0$ can always be written as
\begin{equation}
\pi_{*} \cL_0= \cO_{\cS}(A) \oplus  \cO_{\cS}(B) \oplus \cO_{\cS}(C)
\label{eq:E09}
\end{equation}
for some integers $A,B$ and $C$. Hence, from~\eqref{eq:E03}
\begin{equation}
\pi_{*} \cL_{\gamma}= \cO_{\cS}(A+\gamma) 
\oplus  \cO_{\cS}(B+ \gamma) \oplus \cO_{\cS}(C+\gamma).
\label{eq:E010}
\end{equation}
We must now compute the values of $A,B$ and $C$.
One can show, using a Leray spectral sequence, that for any values of $n$
and
$r$ 
\begin{equation}
H^{0}(\cS, \pi_{*} \cL_{\gamma})= 
H^{0} (\pi^{*}\cS, \cL_{\gamma}).
\label{eq:E08}
\end{equation}
Here, we will continue to assume that $n=3$ and $r=1$.
Now, take $\gamma = 0$. Then the bundle $\cL_{\gamma}$
is equal to $\cL_{0}=\cO_{dP_{9}}(n\sigma|_{dP_{9}})$. Since for $r=1$ one
can
show, using~\eqref{eq:20e} and~\eqref{eq:25}, that
\begin{equation}
\sigma |_{dP_9} \cdot  \sigma |_{dP_9} =-1,
\label{eq:E011}
\end{equation}
the bundle $\cO_{dP_9}(n \sigma |_{dP_9}) $ has only one section and,
hence,
\begin{equation}
h^{0}(\cS, \pi_{*} \cL_{0} )= h^{0} (dP_9, \cL_{0}) = 1,
\label{eq:E011.1}
\end{equation}
where we have used~\eqref{eq:E08}. From equations~\eqref{eq:147} and
~\eqref{eq:148}
we conclude that
\begin{equation}
A=0, \qquad B,C<0.
\label{eq:E012}
\end{equation}
Next, consider the case $\gamma > 2$ that corresponds, 
according to~\eqref{eq:26a}, 
to the case when the spectral cover is chosen to be positive. 
From~\eqref{eq:E28} and~\eqref{eq:E08} we easily find that
\begin{equation}
h^0 (S, \pi_{*} \cL_{\gamma})=3 \gamma - 2 .
\label{eq:E013}
\end{equation}
As in the previous example, one must have
\begin{equation}
\gamma + B \geq 0, \qquad \gamma +C \geq 0,
\label{eq:E014}
\end{equation}
which implies, using~\eqref{eq:147} and~\eqref{eq:E010}, that 
\begin{equation}
h^0 (S, \pi_{*} \cL_{\gamma}) = (\gamma +1) + (\gamma +B +1 )
+(\gamma +C +1). 
\label{eq:E015}
\end{equation}
But, from~\eqref{eq:E013}, this expression must equal $ 3 \gamma -2 $.
Therefore 
\begin{equation}
B+C = -5.
\label{eq:E016}
\end{equation}
To find the values of $B$ and $C$ we use the same techniques as in the case
$z=\cE$. For any divisor ${\cal{D}}$ in $dP_9$ we have the 
following short exact sequence 
\begin{equation}
0 \rightarrow \cO_{dP_9} ({\cal{D}}-\sigma|_{dP_9}) \rightarrow
\cO_{dP_9}({\cal{D}}) \rightarrow \cO_{\sigma|_{dP_9}}({\cal{D}} 
\cdot \sigma|_{dP_9}) \rightarrow 0,
\label{eq:Z1}
\end{equation}
implying the long exact sequence of cohomology groups given by
\begin{eqnarray}
&& 0 \rightarrow H^0 (dP_9, \cO_{dP_9} ({\cal{D}}-\sigma|_{dP_9})) 
\rightarrow H^0 (dP_9, \cO_{dP_9} ({\cal{D}}))
\rightarrow  H^0 ({\mathbb P}^1, \cO_{{\mathbb P}^1} 
({\cal{D}} \cdot\sigma|_{dP_9})) \nonumber \\
&&\rightarrow H^1 (dP_9, \cO_{dP_9} ({\cal{D}} -\sigma|_{dP_9})) 
\rightarrow.
\label{eq:Z1a}
\end{eqnarray}
First, assume that ${\cal{D}} \cdot \sigma|_{dP_9} <
0$. Then, it follows from~\eqref{eq:148} 
that $h^{0}({\mathbb P}^{1}, \cO_{{\mathbb
P}^{1}}({\cal{D}} \cdot \sigma|_{dP_9}))=0$ and, hence
\begin{equation}
h^0 (dP_9, \cO_{dP_9} ({\cal{D}})) = 
h^0 (dP_9, \cO_{dP_9}({\cal{D}}-\sigma|_{dP_9})),
\qquad  {\cal{D}}\cdot \sigma|_{dP_9} < 0. 
\label{eq:Z2} 
\end{equation}
Now assume ${\cal{D}} \cdot \sigma|_{dP_9} \geq 0$.
Equation~\eqref{eq:147} 
then implies 
\begin{equation}
h^{0}({\mathbb P}^{1}, \cO_{{\mathbb P}^{1}}({\cal{D}} \cdot
\sigma|_{dP_9}))=
{\cal{D}} \cdot \sigma|_{dP_9}+1.
\label{eq:Z2a}
\end{equation}
Note, using $\sigma|_{dP_9}^{2}=-1$, that 
$({\cal{D}}-\sigma|_{dP_9}) \cdot \sigma|_{dP_9}
>0$. If we further assume that ${\cal{D}} \cdot F >1$, then $({\cal{D}}
 -\sigma|_{dP_9}) \cdot F >0$ and, hence, ${\cal{D}} -\sigma|_{dP_9}$ 
is a positive divisor in $dP_9$. It
then follows from the Kodaira vanishing theorem that
\begin{equation}
H^{q}(dP_9,\cO_{dP_9}({\cal{D}} -\sigma|_{dP_9}))=0
\label{eq:Z2b}
\end{equation}
for $q > 0$. Hence, the exact sequence~\eqref{eq:Z1a} truncates to 
\begin{equation}
0 \rightarrow H^0 (dP_9, \cO_{dP_9} ({\cal{D}}-\sigma|_{dP_9})) 
\rightarrow H^0 (dP_9, \cO_{dP_9} ({\cal{D}}))
\rightarrow  H^0 ({\mathbb P}^1, \cO_{{\mathbb P}^1} 
({\cal{D}} \cdot\sigma|_{dP_9})) \rightarrow 0. 
\label{eq:Z2c}
\end{equation}
Therefore, using~\eqref{eq:Z2a}, for ${\cal{D}}-\sigma|_{dP_9} > 0$ we
have 
\begin{equation}
h^0 (dP_9, \cO_{dP_9} ({\cal{D}})) - 
h^0 (dP_9, \cO_{dP_9} ({\cal{D}}-\sigma|_{dP_9})) =
{\cal{D}}\cdot \sigma|_{dP_9} +1, 
\qquad  {\cal{D}}  \cdot \sigma|_{dP_9} \geq 0. 
\label{eq:Z3}
\end{equation}
Now, from equations~\eqref{eq:Z2} and~\eqref{eq:Z3} we can compute 
$h^0 (dP_9, \cO_{dP_9} (k \sigma|_{dP_9} +\gamma F))$ for $k=0,1,2,3$.
Start with $k=0$. In this case, equations~\eqref{eq:E08}
and~\eqref{eq:147} 
imply
that
\begin{equation}
h^0(dP_9, \cO_{dP_9} (\gamma F)) = h^0(dP_9, \pi^{*}\cO_{\cS}(\gamma ))=
h^0(\cS, \cO_{\cS}(\gamma)) = \gamma +1, \qquad \gamma \geq 0.
\label{eq:Z5}
\end{equation}
Now take $k=1$. By the same argument as in the previous case, the only freedom
to move a divisor of the class $\sigma|_{dP_9} + \gamma F$ is in the 
fibers, so we conclude
\begin{equation}
h^0 (dP_9, \cO_{dP_9} (\sigma|_{dP_9} +\gamma  F))=
h^0 (dP_9, \cO_{dP_9} (\gamma F))= \gamma +1, \qquad \gamma \geq 0.
\label{eq:Z6}
\end{equation}  
As the next step, take $k=2$ and set ${\cal{D}}=2\sigma|_{dP_9}+\gamma F$. 
First, note that
\begin{equation}
(2\sigma|_{dP_9}+\gamma F) \cdot \sigma|_{dP_9} =\gamma -2.
\label{eq:Z6a}
\end{equation}
This breaks into two cases. For $\gamma=0, 1$, ${\cal{D}} \cdot 
\sigma|_{dP_9}<0$. Then, it follows from~\eqref{eq:Z2} and~\eqref{eq:Z6}
that
\begin{equation}
h^0 (dP_9, \cO_{dP_9} (2\sigma|_{dP_9} +\gamma F))=
h^0 (dP_9, \cO_{dP_9} (\sigma|_{dP_9} +\gamma F)) =\gamma +1, 
\qquad \gamma =0,1 .
\label{eq:Z7}
\end{equation}  
Now let $\gamma  \geq 2$. 
In this case, ${\cal{D}} \cdot \sigma|_{dP_9} \geq 0$ and
\begin{equation}
(\sigma|_{dP_9}+\gamma F) \cdot \sigma|_{dP_9} =\gamma -1 >0, \qquad 
(\sigma|_{dP_9}+\gamma F) \cdot F=1 >0.
\label{eq:Z7a}
\end{equation}
Then equation~\eqref{eq:Z3} is applicable and implies that
\begin{equation}
h^0 (dP_9, \cO_{dP_9} (2\sigma|_{dP_9} +\gamma  F))=
h^0 (dP_9, \cO_{dP_9} (\sigma|_{dP_9} +\gamma F)) +\gamma -1  , 
\qquad   \gamma  \geq 1, 
\label{eq:Z8}
\end{equation}  
where we have used~\eqref{eq:Z6}. Finally, consider $k=3$
and set ${\cal{D}}=3\sigma|_{dP_9}+\gamma F$. Then
\begin{equation}
(3\sigma|_{dP_9}+\gamma F) \cdot \sigma|_{dP-9}=\gamma -3.
\label{eq:Z8a}
\end{equation}
Hence, for $0 \leq \gamma \leq 2$, ${\cal{D}} \cdot \sigma|_{dP_9} < 0$
and
equation~\eqref{eq:Z2} implies
\begin{equation}
h^0 (dP_9, \cO_{dP_9} (3\sigma|_{dP_9} +\gamma F))=
h^0 (dP_9, \cO_{dP_9} (2\sigma|_{dP_9} +\gamma F)) , 
\qquad 0 \leq  \gamma  \leq 2. 
\label{eq:Z9}
\end{equation}  
For $\gamma \geq 3$, ${\cal{D}} \cdot \sigma|_{dP_9} \geq 0$ and 
\begin{equation}
(2\sigma|_{dP_9} +\gamma F) \cdot \sigma|_{dP_9}=\gamma -2 > 0, \qquad 
(2\sigma|_{dP_9} +\gamma F) \cdot F=2 > 0.
\label{eq:Z9a}
\end{equation}
Then, equation~\eqref{eq:Z3} is applicable and gives
\begin{equation}
h^0 (dP_9, \cO_{dP_9} (3\sigma|_{dP_9} +\gamma F))=
h^0 (dP_9, \cO_{dP_9} (2\sigma|_{dP_9} +a\gamma F)) + 
\gamma -2 , \qquad  \gamma  \geq 2. 
\label{eq:Z10}
\end{equation}  
To get the required information about the coefficients
$B$ and $C$, it is sufficient to look at one particular value of $\gamma$,
namely $\gamma=4$. 
In this case, expressions~\eqref{eq:Z8} and~\eqref{eq:Z9} tell us
that
\begin{equation}
h^0 (dP_9, \cO_{dP_9} (3\sigma|_{dP_9} +\gamma F)) = 4.
\label{eq:Z11}
\end{equation}
By using~\eqref{eq:E010}, ~\eqref{eq:E08},~\eqref{eq:E012},
and~\eqref{eq:Z11}, we conclude 
\begin{equation}
h^{0}(\cS, \pi_{*}\cL_{\gamma} )=h^0 (\cS, \cO_{\cS}(2) \oplus
\cO_{\cS}(2+B) \oplus \cO_{\cS}(2+C))=4.
\label{eq:Z12}
\end{equation}
Since the line bundle $\cO_{\cE}(2)$ already has $3$ sections, one of the
other two line bundles has to be trivial and one has to be negative.
Without loss of generality, one can choose
\begin{equation}
B=-2, \qquad C \leq -3
\label{eq:Z12a} 
\end{equation}
as the solution of equation~\eqref{eq:Z12}. Combining this result
with~\eqref{eq:E016}, we see that
\begin{equation}
B = -2, \qquad C = -3.
\label{eq:E017}
\end{equation}
Putting everything together, we conclude that, for $n=3$, $r=1$ and
arbitrary
coefficients $a$ and $b$,    
one has
\begin{equation}
\pi_{*} \cL_{\gamma} = \cO_{\cS} (b-(a+1)) \oplus 
\cO_{\cS} (b-(a+3))  \oplus  \cO_{\cS} (b-(a+4)).
\label{eq:E018}
\end{equation}
The holomorphic sections of this vector bundle, specified by
\begin{equation}
h^{0}(\cS, \pi_{*}\cL_{\gamma})=3(b-(a+1))-2
\label{eq:E018a}
\end{equation}
are in one-to-one correspondence with the small instanton transition
moduli.
It is useful at this point to give a concrete example.\\

\noindent $\bf Example: \rm$ Consider the small instanton transition
presented
in Example 2 earlier in this Section. Recall that the pre-transition
vector
bundle $V$ was specified by $B={\mathbb F}_{1}$, $G=SU(3)$, $a=7$, $b=12$
and
$\lambda=1/2$. In this example, the $M$5-brane involved in the transition
is
wrapped on the curve
\begin{equation}
z=\cS.
\label{eq:E018b}
\end{equation}
Then, from~\eqref{eq:E018} we find that the transition moduli are the
holomorphic
sections of the rank $3$ vector bundle
\begin{equation}
\pi_{*}\cL_{4}=\cO_{\cS}(4) \oplus \cO_{\cS}(2) \oplus \cO_{\cS}(1) 
\label{eq:E018c}
\end{equation}
over $z=\cS$, where, from~\eqref{eq:E02}
\begin{equation}
\cL_{4}=\cO_{dP_{9}}(3\sigma|_{dP_{9}}+4F).
\label{eq:E018d}
\end{equation}
It follows from~\eqref{eq:E018a} that the number of transition moduli is
\begin{equation}
h^{0}(\cS,\pi_{*}\cL_{4})=10,
\label{eq:E018e}
\end{equation}
in agreement with~\eqref{eq:80cc} and~\eqref{eq:E28ee}.\\

The above method of finding the push-forwards of the line bundles
$\cL_{a}$ and $\cL_{\gamma}$ onto their respective curves $z$ is rather
technical. It is useful, therefore, to give a more intuitive, algebraic
derivation of the rank $3$ vector bundles~\eqref{eq:155} and~\eqref{eq:E018}.
To do this, recall from Section 3
that the Calabi-Yau threefold $X$ is given by the 
Weierstrass equation~\eqref{eq:1}, with $x$, $y$ and $z$ being sections 
of the following line bundles
\begin{equation}
x \sim \cO_{P}(1) \otimes \cL^2 , \qquad 
y \sim \cO_{P}(1) \otimes \cL^3 , \qquad
z \sim \cO_{P}(1).
\label{eq:new1}
\end{equation}
First consider $z=\cE$ and let $P^{\prime}$ denote the restriction of the 
${\mathbb C}{\mathbb P}^2$-bundle $P$
to the curve $\cE$. Then the elliptically fibered surface 
$\pi^{*}\cE = K3$ is defined by a similar Weierstrass equation,
but now as a divisor in $P^{\prime}$. We find the following identifications
\begin{equation}
\cL|_{K3} = \cO_{K3} (2F) , \qquad 
\cO_{P} (1)|_{K3} = \cO_{K3} (3 \sigma|_{K3}).
\label{eq:new2}
\end{equation}
The first identification follows from the definition of $\cL$ 
as the pullback
to $P$ of the conormal bundle to 
$\sigma(B)$ in $X$. To see this, note that when we restrict $\cL$ to $\sigma|_{K3}$,
we get the conormal bundle to $\sigma|_{K3}$ in $K3$, which is 
$\cO_{\sigma|_{K3}} (2)$. Pulling this bundle back to $K3$ gives
$\cO_{K3} (2F)$. The second identification follows from  
the fact that $\cO_{P} (1) = \cO_X (3 \sigma)$. 
Using equations~\eqref{eq:new1} and~\eqref{eq:new2}, we see from the
Weierstrass equation defining $K3$ that $x, y$ and $z$ are sections of 
the line bundles
\begin{equation}
x \sim \cO_{K3}(3 \sigma|_{K3} + 4F) , \qquad 
y \sim \cO_{K3}(3 \sigma|_{K3} + 6F) , \qquad
z \sim \cO_{K3}(3 \sigma|_{K3} ).
\label{eq:new3}
\end{equation}
Recall that our problem is to find the push-forward of the line 
bundle $\cL_{a}=\cO_{K3}(3 \sigma|_{K3} + aF)$. Denote by $s_k$ a section
of $H^0 (\cE, \cO_{\cE}(k))$. Using a Leray spectral sequence, one can show
that
\begin{equation}
H^0 (\cE, \cO_{\cE}(k))=H^0 (K3, \cO_{K3}(kF)).
\label{eq:new3a}
\end{equation}
Hence, $s_{k}$ lifts to a section of $H^0 (K3, \cO_{K3}(kF))$, which we will
also denote by $s_{k}$. Clearly, one can construct the following sections
of $\cL_a$
\begin{equation}
s_a z+ s_{a-4} x + s_{a-6} y.
\label{eq:new4}
\end{equation}
This shows explicitly how sections of $H^0 (K3, \cL_a)$ can be written 
as a sum of three terms involving sections of 
$H^0 (\cE, \cO_{\cE}(a))$, $H^0 (\cE, \cO_{\cE}(a-4))$ and  
$H^0 (\cE, \cO_{\cE}(a-6))$ respectively. We conclude that
\begin{equation}
\pi_{*}\cL_{a}=\cO_{\cE}(a) \oplus \cO_{\cE}(a-4) \oplus \cO_{\cE}(a-6), 
\label{eq:new4a}
\end{equation}
exactly as in~\eqref{eq:155}.

A similar argument applies to the case 
$z=\cS$. Let $P^{\prime \prime}$ denote the restriction of the bundle $P$
to the curve $\cS$. The Weierstrass equation~\eqref{eq:1} now defines 
the elliptic surface $dP_9$ as a divisor in $P^{\prime \prime}$. 
Since $\sigma|_{dP_9} \cdot \sigma|_{dP_9} =-1$, the conormal bundle 
to $\sigma|_{dP_9}$ in $dP_9$ is $\cO_{\sigma|_{dP_9}} (1)$. As a result
\begin{equation}
\cL|_{dP_9} = \cO_{dP_9} (F) , \qquad 
\cO_{P} (1)|_{dP_9} = \cO_{dP_9} (3 \sigma|_{dP_9}).
\label{eq:new5}
\end{equation}
Using this and  equation~\eqref{eq:new1}, we find 
that $x, y$ and $z$ are now sections of the following line
bundles on $dP_9$
\begin{equation}
x \sim \cO_{dP_9}(3 \sigma|_{dP_9} + 2F) , \qquad 
y \sim \cO_{dP_9}(3 \sigma|_{dP_9} + 3F) , \qquad
z \sim \cO_{dP_9}(3 \sigma|_{dP_9} ).
\label{eq:new6}
\end{equation}
If $s_k$ denotes a section of $H^0 (\cS, \cO_{\cS}(k))= H^0 (dP_{9},
\cO_{dP_{9}}(kF))$, we can construct 
the following sections of the line bundle 
$\cL_{\gamma} = \cO_{dP_9} ( 3 \sigma|_{dP_9} + \gamma F )$
\begin{equation}
s_{\gamma} z + s_{\gamma -2} x +s_{\gamma -3} y.
\label{eq:new7}
\end{equation}
 This shows 
how sections of $H^0 (dP_9, \cL_{\gamma})$ can be written as a 
sum of three terms involving sections of 
$H^0 (\cS, \cO_{\cS}(\gamma))$, $H^0 (\cS, \cO_{\cS}(\gamma -2))$ and  
$H^0 (\cS, \cO_{\cS}(\gamma -3))$ respectively. We conclude that
\begin{equation}
\pi_{*}\cL_{\gamma}=\cO_{\cS}(\gamma) \oplus \cO_{\cS}(\gamma-2) 
\oplus \cO_{\cS}(\gamma-3),
\label{eq:new7a}
\end{equation}
exactly as in~\eqref{eq:E018}.

\subsection*{Acknowledgements:}

We would like to thank Tony Pantev and Rene Reinbacher for helpful
conversations. Evgeny Buchbinder and Burt Ovrut are supported in part 
by the DOE under contract No. DE-AC02-76-ER-03071. Ron Donagi is supported 
in part by an NSF grant DMS-9802456.


\end{document}